\documentclass[a4paper,11pt]{article}
\pdfoutput=1 

\usepackage{jheppub} 

\usepackage[T1]{fontenc} 

\title{Consistent QFT description of non-standard neutrino interactions}

\usepackage{graphicx}
 \usepackage{dcolumn}
\usepackage{bm}
\usepackage{color}
\usepackage[normalem]{ulem}
\usepackage{hyperref}
\usepackage{braket} 
\usepackage{amsmath}
\usepackage{cancel}

\begin{document}
\newcommand{\fref}[1]{Fig.~\ref{fig:#1}} 
\newcommand{\eref}[1]{Eq.~\eqref{eq:#1}} 
\newcommand{\erefn}[1]{ (\ref{eq:#1})}
\newcommand{\erefs}[2]{Eqs.~(\ref{eq:#1}) - (\ref{eq:#2}) } 
\newcommand{\aref}[1]{Appendix~\ref{app:#1}}
\newcommand{\sref}[1]{Section~\ref{sec:#1}}
\newcommand{\cref}[1]{Chapter~\ref{ch:.#1}}
\newcommand{\tref}[1]{Table~\ref{tab:#1}}

\newcommand{\nn}{\nonumber \\}  
\newcommand{\nnl}{\nonumber \\}  
\newcommand{\nl}{& \nonumber \\ &}
\newcommand{\bnl}{\right .  \nonumber \\  \left .}
\newcommand{\dbnl}{\right .\right . & \nonumber \\ & \left .\left .}

\newcommand{\beq}{\begin{equation}} 
\newcommand{\eeq}{\end{equation}} 
\newcommand{\ba}{\begin{array}}  
\newcommand{\ea}{\end{array}} 
\newcommand{\bea}{\begin{eqnarray}}  
\newcommand{\eea}{\end{eqnarray} }  
\newcommand{\be}{\begin{eqnarray}}  
\newcommand{\ee}{\end{eqnarray} }  
\newcommand{\bal}{\begin{align}}
\newcommand{\eal}{\end{align}}   
\newcommand{\bi}{\begin{itemize}}  
\newcommand{\ei}{\end{itemize}}  
\newcommand{\ben}{\begin{enumerate}}  
\newcommand{\een}{\end{enumerate}}  
\newcommand{\bc}{\begin{center}}
\newcommand{\ec}{\end{center}} 
\newcommand{\bt}{\begin{table}}
\newcommand{\et}{\end{table}}  
\newcommand{\btb}{\begin{tabular}}
\newcommand{\etb}{\end{tabular}}  
\newcommand{\bvec}{\left ( \ba{c}}
\newcommand{\evec}{\ea \right )}

\newcommand{\cO}{{\mathcal O}} 
\newcommand{\co}{{\mathcal O}} 
\newcommand{\cL}{{\mathcal L}} 
\newcommand{\cl}{{\mathcal L}} 
\newcommand{\cM}{{\mathcal M}}

\newcommand{\const}{\mathrm{const}}

\newcommand{\ev}{ \mathrm{eV}}
\newcommand{\kev}{\mathrm{keV}}
\newcommand{\mev}{\mathrm{MeV}}
\newcommand{\gev}{\mathrm{GeV}}
\newcommand{\tev}{\mathrm{TeV}}

\newcommand{\mpl}{M_{\mathrm Pl}}

\def\mgut{\, M_{\rm GUT}}
\def\tgut{\, t_{\rm GUT}}
\def\mpl{\, M_{\rm Pl}}
\def\mkk{\, M_{\rm KK}}
\newcommand{\msusy}{M_{\rm soft}}

\newcommand{\dslash}[1]{#1 \! \! \! {\bf /}}
\newcommand{\ddslash}[1]{#1 \! \! \! \!  {\bf /}}

\def\ads{AdS$_5$\,}
\def\adse{AdS$_5$}
\def\intdk{\int {d^4 k \over (2 \pi)^4}} 

\def\ra{\rangle}
\def\la{\langle}  

\def\sgn{{\rm sgn}}
\def\pa{\partial}  
\newcommand{\dlr}{\overleftrightarrow{\partial}}
\newcommand{\Dlr}{\overleftrightarrow{D}}
\newcommand{\re}{{\mathrm{Re}} \,}
\newcommand{\im}{{\mathrm{Im}} \,}
\newcommand{\tr}{\mathrm T \mathrm r}  

\newcommand{\Ra}{\Rightarrow}
\newcommand{\lra}{\leftrightarrow}
\newcommand{\llra}{\longleftrightarrow}

\newcommand\simlt{\stackrel{<}{{}_\sim}}
\newcommand\simgt{\stackrel{>}{{}_\sim}}   
\newcommand{\zt}{$\mathbb Z_2$ }

\newcommand{\ha}{{\hat a}}
\newcommand{\hab}{{\hat b}}
\newcommand{\hac}{{\hat c}} 

\newcommand{\ti}{\tilde}  
\def\hc{{\rm h.c.}} 
\def\ov{\overline}  
  

\newcommand{\eps}{\epsilon}
\newcommand{\eS}{\epsilon_S}
\newcommand{\eT}{\epsilon_T}
\newcommand{\eP}{\epsilon_P}
\newcommand{\eL}{\epsilon_L}
\newcommand{\eR}{\epsilon_R}
\newcommand{\teps}{{\tilde{\epsilon}}}
\newcommand{\teS}{{\tilde{\epsilon}_S}}
\newcommand{\teT}{{\tilde{\epsilon}_T}}
\newcommand{\teP}{{\tilde{\epsilon}_P}}
\newcommand{\teL}{{\tilde{\epsilon}_L}}
\newcommand{\teR}{{\tilde{\epsilon}_R}}
\newcommand{\eLc}{{\epsilon_L^{(c)}}}
\newcommand{\eLv}{{\epsilon_L^{(v)}}}
\newcommand{\eSP}{\epsilon_{S,P}}
\newcommand{\teSP}{{\tilde{\epsilon}_{S,P}}}

\newcommand{\lz}{\lambda_z}
\newcommand{\dgz}{\delta g_{1,z}}
\newcommand{\dkg}{\delta \kappa_\gamma}

\def\cog{\color{OliveGreen}}
\def\cor{\color{Red}}
\def\copu{\color{purple}}
\def\coro{\color{RedOrange}}
\def\coma{\color{Maroon}}
\def\cob{\color{Blue}}
\def\cobr{\color{Brown}}
\def\cobl{\color{Black}}
\def\cost{\color{WildStrawberry}}

\newcommand{\tl}{{\tilde{\lambda}}}
\newcommand{\dll}{{\frac{\delta\lambda}{\lambda}}}

\preprint{IFIC/19-39, FTUV/19-1007, LPT-Orsay-19-35}

\author[a]{Adam Falkowski,}
\author[b]{Mart\'{i}n Gonz\'{a}lez-Alonso,}
\author[c,d]{Zahra~Tabrizi}


\affiliation[a]{Universit\'{e} Paris-Saclay, CNRS/IN2P3, IJCLab, 91405 Orsay, France}
\affiliation[b]{Departament de F\'isica Te\`orica, IFIC, Universitat de Val\`encia - CSIC, Apt.  Correus 22085, E-46071 Val\`encia, Spain}
\affiliation[c]{Instituto de F\'iısica Gleb Wataghin, Universidade Estadual de Campinas (UNICAMP), Rua S\'ergio Buarque de Holanda, 777, Campinas, SP, 13083-859, Brazil}
\affiliation[d]{Center for Neutrino Physics, Department of Physics, Virginia Tech, Blacksburg, VA 24061, USA}

\abstract{Neutrino oscillations are precision probes of new physics.
Apart from neutrino masses and mixings, they are also sensitive to possible deviations of low-energy interactions between quarks and leptons from the Standard Model predictions. 
In this paper we develop a systematic description of  such non-standard interactions (NSI) in oscillation experiments within the quantum field theory framework.  
We calculate the event rate and oscillation probability in the presence of general NSI,  starting from the effective field theory (EFT) in which new physics modifies the flavor or Lorentz structure of charged-current interactions between leptons and quarks.
We also provide the matching between the EFT Wilson coefficients and the widely used simplified quantum-mechanical approach, where new physics is encoded in  a set of production and detection NSI parameters. 
Finally, we discuss the consistency  conditions for the standard NSI approach to correctly reproduce the quantum field theory result.}

\maketitle

\section{Introduction} 
Precision measurements at low energies are sensitive probes of fundamental interactions that complement collider searches.
Neutrino oscillation experiments \cite{Bilenky:1978nj} are a specific class thereof where one observes a characteristic oscillatory dependence of the neutrino detection rate as a function of the neutrino energy $E_\nu$ and the distance between the neutrino source and detector. 
The large body of oscillation data so far has established the existence of at least three distinct neutrino states with different masses~\cite{deSalas:2017kay,Esteban:2018azc}, which is consistent with the  predictions of the Standard Model (SM) supplemented with dimension-5 terms leading to Majorana masses for the SM neutrinos~\cite{Weinberg:1979sa}.   
Within this paradigm, neutrino mass squared differences and  angles of the PMNS mixing matrix have been measured with good accuracy. 
This opens the door to also probing and constraining new physics (NP), by which we mean non-standard interactions (NSI) between neutrinos and matter that arise from physics beyond the SM (BSM)~\cite{Bergmann:1999rz,Antusch:2006vwa,Kopp:2007ne,Bolanos:2008km,Ohlsson:2008gx,Delepine:2009am,Biggio:2009nt,Leitner:2011aa,Ohlsson:2012kf,Esmaili:2013fva,Li:2014mlo,Agarwalla:2014bsa,Blennow:2015nxa,Coloma:2017ncl,Farzan:2017xzy,Choudhury:2018xsm,Heeck:2018nzc,Altmannshofer:2018xyo,AristizabalSierra:2018eqm,Esteban:2018ppq}. 
To this end, however, one needs a map between fundamental BSM parameters and observables in oscillation experiments.
In this paper we construct such a map for the EFT of SM degrees of freedom, in which NP modifies the charged-current interactions between neutrinos, charged leptons, and quarks. 
This map makes possible to understand the BSM implications of a given neutrino measurement, but also to combine and compare the obtained bounds with other probes that are sensitive to the same non-standard interactions (such as collider searches, $\mu\to e\gamma$, etc.).

We also discuss the consistency conditions for the widely used quantum-mechanical (QM) approach, where  New Physics is parametrized by a set of NSI production and detection parameters, to correctly reproduce the quantum field theory (QFT) result.
    
The paper has the following structure. 
In Section~\ref{sec:Formalism}, we discuss how to calculate the differential rate of neutrino oscillation events, starting with a QFT with an arbitrary field content and interaction Lagrangian.
The result is encompassed in \eref{rateqft}, where the rate  is expressed by the amplitudes for production and detection of the neutrino.  
This can be compared with the QM prescription to calculate the same observable, in which the connection with the underlying physics of neutrino interactions is obscure. 
In Section~\ref{sec:QFTandNSI}, we introduce an EFT Lagrangian that describes interactions between neutrinos, leptons, and quarks. 
The possible departure of these interactions from the SM predictions is parameterized by a set of Wilson coefficients. 
The connection between these Wilson coefficients and the rate formula  in \eref{rateqft} is made transparent by introducing  process-dependent production and detection coefficients. 
We also derive the matching between the Wilson coefficients in the EFT, and the familiar NSI parameters in the QM  description. 
The matching is always possible at the linear order in the Wilson coefficients. 
However, only if the production and detection coefficients satisfy a certain consistency condition, that matching is valid beyond the linear order. 
In Section~\ref{sec:matching} we calculate the production and detection coefficients for several specific processes of relevance to current neutrino experiments. 
We cover the cases of neutrinos produced in nuclear beta or pion decays, and detected by inverse beta decay. 
The formulas for the neutrino oscillation probability are collected in Section~\ref{sec:oscprob}.  
Section~\ref{sec:Muondecay} is devoted to the discussion of NSI for neutrinos produced in muon decay. The study of this process requires the introduction of 4-lepton effective operators and introduces new features due to the presence of one neutrino and one antineutrino.
Finally, Section~\ref{sec:conclusion} contains our concluding remarks.

 \section{Formalism}
 \label{sec:Formalism}
 
\subsection{QFT description} 
Oscillation probability can be rigorously derived in the framework of quantum field theory. Various derivations are available in the literature in the absence of NSI (see e.g.~\cite{Giunti:1993se,Akhmedov:2010ms,Kobach:2017osm}). 
Below we give an expression valid for completely general interactions between neutrinos and matter. 
Consider neutrinos produced in the process  $S \to X_\alpha \nu$  (e.g. beta decay of a nucleus in a reactor, or pion decay), where $X_\alpha$ is one or more body final states  containing one charged lepton $\ell_\alpha = (e,\mu,\tau)$. 
Neutrinos are detected via the process $\nu\, T \to Y_\beta$ (e.g. inverse beta decay),  where again $Y_\beta$ contains a charged lepton $\ell_\beta$. 
The production and detection can be described by QFT amplitudes  $\cM_{\alpha k}^P \equiv \cM(S \to X_\alpha \nu_k)$  and $\cM_{\beta k}^D \equiv \cM(\nu_k T \to Y_\beta)$, where the index $k$ labels neutrino mass eigenstates. 
The information about fundamental parameters, is encoded in $\cM_{\alpha k}^P$ and $\cM_{\beta k}^D$, 
which should be then connected to observables.
For  source (S) and target (T)  states separated by a macroscopic distance $L$,  the observable is the differential rate of detected events per target particle $R_{\alpha \beta} \equiv {d N_{\alpha \beta}\over N_T dt d E_\nu}$ given by  
\beq
\label{eq:rateqft}
R_{\alpha \beta} =  {N_S \over 32 \pi L^2 m_S m_T E_\nu }  
\sum_{k,l} 
\,e^{-i {L \Delta m_{kl}^2 \over 2 E_\nu}}
\!\int\!  d \Pi_{P'}
\cM_{\alpha k}^P  \bar \cM_{\alpha l}^{P} 
\!\int\!  d \Pi_D \cM_{\beta k}^D   \bar \cM_{\beta l}^{D} .  
\eeq
A compact derivation of this formula is presented in~\aref{oscqft}, where we also enumerate its limitations.  
Above, $\Delta m_{kl}^2 \equiv m_k^2 - m_l^2$ is the mass squared difference between  neutrino eigenstates.
 The phase space elements  $d \Pi_{P}$ and $d \Pi_{D}$ for the production and detection processes are defined in the standard way: 
{$d \Pi \equiv {d^3 k_1 \over (2 \pi)^3 2 E_1} \dots  {d^3 k_n \over (2 \pi)^3 2 E_n} (2\pi)^4 \delta^4(\mathcal{P} - \sum k_i )$}, 
where $\mathcal{P}$ is the total 4-momentum of the initial state and $k_i$ are the 4-momenta of the final states. 
The production $d \Pi_P$ includes the neutrino phase space  ${d^3 k_\nu \over (2 \pi)^3 2 E_\nu }$ and  we define $d \Pi_{P'}$ via  $d \Pi_P \equiv  d \Pi_{P'} dE_\nu$.  
The amplitudes $\cM_{P,D}$ describe the neutrino production and detection processes, which we allow to be arbitrary.
The $\int$ sign in \eref{rateqft} involves integration and sum/average over all unobserved degrees of freedom, such as angular variables and spins. 
Finally, complex conjugated amplitudes are denoted with a bar,  $N_{S,T}$ are the number of source/target particles, and
$m_{S,T}$ are their masses.
The derivation of \eref{rateqft} assumes that neutrinos are produced by a source at rest and are emitted isotropically;
see \eref{NOF_dPdt_generalized} for a more general formula.

The rate in \eref{rateqft} displays the famous oscillatory behavior via the  $\exp\left(-i {L \Delta m_{kl}^2 \over 2 E_\nu}\right)$ factor.  
In the absence of oscillations, rates would be calculated using the  neutrino differential flux $\Phi_\alpha \equiv {N_S \over 4 \pi L^2} {d \Gamma^P_\alpha \over d E_\nu}$, and the detection cross section at the target.
The source decay rate $\Gamma^P_\alpha$ (with  emission of $\ell_\alpha$ and summed over neutrino mass eigenstates) and the detection cross section $\sigma^D_\beta$ (with emission of $\ell_\beta$ and summed over neutrino mass eigenstates) can be calculated by the usual means in QFT.
We have 
\beq
\label{eq:ratenoscQFT}
\Phi_\alpha \sigma_\beta
 = 
{N_S \over 32 \pi L^2 m_S m_T E_\nu }
\int  d \Pi_{P'}    \sum_{k}|\cM_{\alpha k}^P|^2 \int d \Pi_D \sum_{l}  |\cM_{\beta l}^D|^2. 
\eeq
One can define the $\nu_\alpha \to \nu_\beta$ {\em oscillation probability} as the ratio of the rate of detected events in \eref{rateqft} to the no-oscillation expression in \eref{ratenoscQFT}, finding
\begin{small}\beq
\label{eq:oscqft}
P_{\alpha \beta} = 
  {\sum_{k,l}
  e^{-i {L \Delta m_{kl}^2 \over 2 E_\nu}}
  \int  d \Pi_{P'}  \cM_{\alpha k}^P  \bar \cM_{\alpha l}^{P}   
 \int d \Pi_D \cM_{\beta k}^D   \bar \cM_{\beta l}^{D} 
 \over  
\int  d \Pi_{P'}  
    \sum_{k}  |\cM_{\alpha k}^P|^2   \int d \Pi_D  \sum_{l}  |\cM_{\beta l}^D|^2  } . 
\eeq\end{small}
This formula appears in Ref.~\cite{Giunti:2007ry} in a slightly different form without explicit phase space integration.
Oscillation probability is an intuitive and widely employed concept, however strictly speaking  $P_{\alpha \beta}$ is not an observable.  
For this reason in this paper we work mainly with the rate in \eref{rateqft}. Nonetheless, the oscillation probability is also discussed in~\sref{oscprob}.

\subsection{ QM-NSI description} 
Neutrino oscillations are often described in a simple quantum mechanical setting.
One defines flavor states  
as linear combinations of mass eigenstates: 
$\ket{\nu_\alpha} = \sum_k U_{\alpha k} \ket{\nu_k}$, where $U$ is the unitary PMNS mixing matrix. 
In this language,  NSI effects are encoded in parameters $\epsilon_{\alpha \beta}^{s,d}$~\cite{Grossman:1995wx,GonzalezGarcia:2001mp,Farzan:2017xzy}, which
  are defined by the mismatch between pure flavor states and  neutrino states produced at the source and detected at the target, namely~\cite{Ohlsson:2008gx}:
\beq 
\label{eq:NSIdefinition}
|\nu_\alpha^s\rangle =  
\frac{(1 + \epsilon^s)_{\alpha \gamma}}{N_\alpha^s}   |\nu_\gamma \rangle~, 
\ \
	\langle \nu^d_\beta | =     
	\langle \nu_\gamma | \frac{(1 + \epsilon^d)_{\gamma \beta}}{N_\beta^d} ~,	\eeq 
with the normalization
$N_\alpha^s = \sqrt{ [  (1 +\epsilon^s)(1+\epsilon^{s \, \dagger}) ]_{\alpha\alpha}}$, 
$N_\beta^{d} = \sqrt{ [  (1 +\epsilon^{d \, \dagger})(1+\epsilon^{d}) ]_{\beta\beta}}$. 
The probability of $\ket{\nu_\alpha^s}$ oscillating into  $\ket{\nu_\beta^d}$ is given by
$P^{\scriptscriptstyle{QM}}_{\alpha \beta} = |\bra{\nu_\beta^d}e^{-iHL}\ket{\nu^s_\alpha}|^2$, 
where the Hamiltonian is 
$H_{\beta \alpha}= \sum_{k}U_{\beta k} m_k^2 U^*_{\alpha k}/(2E_\nu)$ in the absence of  matter effects in propagation. 
In this approach, which we refer to as {\em QM-NSI formalism}, the event rate is given by~\cite{Antusch:2006vwa}
\bea
\label{eq:ratensi}
R_{\alpha \beta}^{\rm QM} 
&=& 
\Phi^{\rm SM}_\alpha \sigma^{\rm SM}_\beta 
P_{\alpha \beta}^{\rm QM} 
(N_\alpha^s N_\beta^d)^2\\
&=&
\Phi^{\rm SM}_\alpha \sigma^{\rm SM}_\beta
\sum_{k,l}  
 e^{-i {L \Delta m_{kl}^2 \over 2 E_\nu}}
  [x_s]_{\alpha k} 
  [x_s]^*_{\alpha l}
  [x_d]_{\beta k} 
  [x_d]^*_{\beta l}~,\nonumber
\eea 
where $x_s \equiv (1 + \epsilon^s)U^* $ and $x_d \equiv (1 + \epsilon^d)^T U$. 
For antineutrinos~\eref{ratensi} holds with $U \to U^*$.
Above, $\Phi^{\rm SM}_\alpha$ and $\sigma^{\rm SM}_\beta$ are the incident flux and detection cross section calculated in the absence of NSI.  
Normalization factors $N_\alpha^s N_\beta^d$ cancel in the observable rate and thus one could have omitted them altogether~\cite{Antusch:2006vwa}; their only role is to ensure that $P_{\alpha \beta} \leq 1$, that is it can be interpreted as a probability.

Results from oscillation experiments are often presented or recast as constraints on the NSI parameters $\epsilon_{\alpha \beta}^{s,d}$. 
However, the utility of the latter hinges on whether they can be unambiguously connected to more fundamental parameters in a Lagrangian.
Only after such matching the coefficients $\epsilon_{\alpha \beta}^{s,d}$ determined in different experimental settings can be meaningfully compared and combined. 
In the following we discuss this issue, and illustrate it with physically relevant examples. 
We will define the conditions under which the NSI parameters can indeed provide an adequate description of NP effects in neutrino oscillation. 
Conversely, we will show examples where this is not the case.  

\section{Matching QFT and QM-NSI results}
\label{sec:QFTandNSI}
%
One could try to match the QM-NSI and QFT language starting from the definition in \eref{NSIdefinition}. 
This however would be problematic, as such concepts as neutrino flavor states or production and detection states are murky in a QFT framework when general charged-current neutrino interactions with matter are allowed. 
Therefore, we will follow a pragmatic approach and match the {\em observable rates} predicted by the QFT (\eref{rateqft}) and QM-NSI frameworks (\eref{ratensi}). 
This comparison  will allow us to determine the map between the NSI  $\epsilon^{s,d}$ and the Lagrangian parameters, or else conclude the map does not exist. 

Thus, in this paper we focus on NP in charged-current interactions between neutrino and matter. 
The theory framework we consider is the EFT of the SM degrees of freedom at the energy scale $\mu \approx 2$~GeV, in which lepton-number conserving NP modifies the effective 4-fermion interactions between leptons and quarks. Extensions to other theories and interactions are straightforward using our approach.
For example, introducing  right-handed neutrinos interacting with matter or additional sterile neutrinos mixing with the active ones, would not bring any qualitative change to the formalism.

At leading order in our EFT  neutrino interactions with matter can be parametrized by the Lagrangian (see e.g. Ref.~\cite{Cirigliano:2012ab})
\bea 
\label{eq:EFT_lweft}
{\cal L}
& \supset &
- \,\frac{2 V_{ud}}{v^2} \big \{   
\left [ {\bf 1} +    \epsilon_L \right]_{\alpha \beta}  (\bar u  \gamma^\mu P_L d)  (\bar \ell_\alpha  \gamma_\mu P_L \nu_\beta) 
\nnl && ~~~+  
[\epsilon_R]_{\alpha \beta}  (\bar u  \gamma^\mu P_R d)   (\bar \ell_\alpha  \gamma_\mu P_L \nu_\beta)
\nnl &&
~~~+\, {1 \over 2 } [\epsilon_S]_{\alpha \beta} (\bar u d) (\bar \ell_\alpha P_L  \nu_\beta)  
- {1 \over 2} [\epsilon_P]_{\alpha \beta} (\bar u \gamma_5 d) (\bar \ell_\alpha P_L  \nu_\beta)   
\nnl 
&&
~~~+\,  {1 \over 4} [\epsilon_T]_{\alpha \beta} (\bar u  \sigma^{\mu \nu} P_L d)   (\bar \ell_\alpha  \sigma_{\mu \nu} P_L \nu_\beta) 
+ \hc  \big \}~,
\eea
where $v \equiv (\sqrt{2} G_F)^{-1/2} \approx 246$~GeV, $V_{ud}$ is a CKM matrix element, 
$\sigma^{\mu\nu}=i[\gamma^\mu,\gamma^\nu]/2$, and $P_{L,R}=(1\mp\gamma_5)/2$.
The quarks $u$, $d$, and charged leptons $\ell_\alpha$  are in the basis where their kinetic and mass terms are diagonal. For  neutrino fields  kinetic terms are diagonal but mass terms are not, thus $\nu_\alpha$ are connected to  mass eigenstates by the unitary PMNS rotation $\nu_{\alpha} = \sum_k U_{\alpha k} \nu_k$. 
In this EFT the effects of NP are parametrized by the Wilson coefficients $[\epsilon_X]_{\alpha \beta}$, which encode new interactions between quarks and leptons mediated by BSM particles heavier than $\sim 2$~GeV. 
For example, non-zero $\epsilon_R$ can arise in left-right symmetric models due to the $W_R$ boson  coupling to right-handed quarks and mixing with the SM $W^\pm$, while non-zero $\epsilon_{S,P,T}$ are generally predicted in leptoquarks models. 
More generally, $\epsilon_X$ can be connected to parameters of the weak-scale EFT, known as the SMEFT~\cite{Cirigliano:2012ab,Falkowski:2019xoe,Bischer:2019ttk}.  
The constraints obtained from neutrino oscillations may have an impact on these broad classes of models. 

We remark that, in the EFT below the weak scale where charged and neutral leptons are not collected into doublets, nothing distinguishes  the basis of $\nu_\alpha$ in \eref{EFT_lweft} as soon as $[\epsilon_L]_{\alpha \beta} \neq 0$.
Unitary rotations $\nu_\alpha \to V \nu_\alpha$ transform the Lagrangian into an equivalent basis with NP parameters rotated as $\delta_{X,L} + \epsilon_X \to (\delta_{X,L} + \epsilon_X) V$, and the neutrino mass matrix rotated by $M_\nu \to V^T M_\nu V$.
Physics of course cannot depend on which basis we work with. 
We will see that observable rates will be invariant under such rotations of $\epsilon_X$ accompanied by rotating the PMNS matrix $U \to V^\dagger U$.

\subsection{ SM interactions} 
To warm up, let us first calculate the event rate in the limit of the SM interactions, which corresponds to setting all  $\epsilon_X = 0$.   In this case, which was studied in Ref.~\cite{Giunti:1993se}, the amplitudes can be decomposed as:
 \beq
\cM_{\alpha k}^{P} = U^{*}_{\alpha k} A^{P}_L~,
\quad\quad
\cM_{\alpha k}^{D} = U_{\alpha k} A^{D}_L~.
\eeq
The functions $A^{P,D}_L$ are independent of the neutrino mass index $k$ up to  negligible corrections, whereas they do depend on the charged-lepton flavor index $\alpha$ (which we omit to ease the notation). They also 
 depend on the kinematic and spin variables in the production and detection  processes, and they appear in the observables  integrated/averaged over  by  $\int d \Pi_{P',D}$. 
All in all the rate in \eref{rateqft} can be written as 
\beq
R^{\rm SM}_{\alpha \beta} =  
\Phi^{\rm SM}_\alpha \sigma^{\rm SM}_\beta 
 \sum_{k,l}  e^{-i {L \Delta m_{kl}^2 \over 2 E_\nu} }
 U^*_{ \alpha k} U_{\alpha l}  U_{ \beta k} U_{\beta l}^*,
\eeq
where the SM flux and cross-section are given by
\beq
\label{eq:SMflux}
\Phi^{\rm SM}_\alpha = \frac{N_S\!\int \!d \Pi_{P'}  |A_L^P|^2}{8 m_S \pi L^2},\quad
\sigma^{\rm SM}_\beta = \frac{ \int\!  d \Pi_D   |A^D_L|^2 }{ 4 E_\nu m_T}. 
\eeq
Exactly the same  result is obtained  from \eref{ratensi} in the limit $\epsilon_{\alpha \beta}^{s,d} = 0$.
   
\subsection{ $V$-$A$ interactions} 
A less trivial example is  when NP enters only via $V$-$A$ interactions: $[\epsilon_L]_{\alpha \beta} \neq 0$~\cite{Bergmann:1999rz,Antusch:2006vwa,Khan:2013hva,Blennow:2016jkn}.
In this case the detection/production amplitudes decompose as 
 \beq
\cM_{\alpha k}^P = [(1 + \epsilon_L)U]^*_{\alpha k} A^P_L~,
\quad\quad
\cM_{\alpha k}^D = [(1 + \epsilon_L)U]_{\alpha k} A^D_L~.
\eeq
We obtain  
\beq  
\label{eq:rateVmA}
R^{V\!-\!A}_{\alpha \beta} = 
\Phi^{\rm {\scriptscriptstyle SM}}_\alpha \sigma^{\rm {\scriptscriptstyle SM}}_\beta
 \sum\limits_{k,l}   
 e^{-i {L \Delta m_{kl}^2 \over 2 E_\nu}}
  [x_L]_{\alpha k}^*   [x_L]_{\alpha l} 
  [x_L]_{\beta k} [x_L]_{\beta l}^* ~,
\eeq 
where $x_L\equiv (1+\epsilon_L)\,U$. 
In fact, the quantity $x_L$ is equivalent to a ``non-unitary mixing matrix", an approach that has been studied in the neutrino literature~\cite{Antusch:2006vwa,Blennow:2016jkn,Xing:2011ur,Cates:2011pz}.
The same result is obtained in the QM-NSI approach from \eref{ratensi} when  NSI parameters are mapped to the Lagrangian parameters as
~\cite{Blennow:2016jkn}:
\beq
{\rm \bf V\!\!-\!\!A:} \quad\quad 	
\epsilon^s_{\alpha \beta}  = [\epsilon_L]^*_{\alpha \beta}, 
\quad \quad
\epsilon^d_{\beta\alpha} = [\epsilon_L]_{\alpha \beta } .
\eeq 
In the $V$-$A$ case the map between NSI and Lagrangian parameters is well-defined, unambiguous, and simple. 

\subsection{General case}
For general NP interactions in \eref{EFT_lweft}, the production and detection amplitudes can be decomposed as
\bea
\label{eq:decomposition}
\cM_{\alpha k}^P &=& U_{\alpha k}^{*} A^P_L + \sum_X [\epsilon_X U]_{\alpha k}^*A^P_X,
\nnl 
\cM_{\beta k}^D &= &  U_{\beta k} A^D_L +  \sum_X [\epsilon_X U]_{\beta k} A^D_X . 
\eea
The sum above goes over all types of interactions in \eref{EFT_lweft}: $X = L,R,S,P,T$. We stress that  $A^{P,D}_X$ will typically have completely different dependence on kinematic and spin variables for different $X$. 
Plugging this decomposition into \eref{rateqft} we obtain
\begin{small}
\bea
\label{eq:rategeneraleft}
R_{\alpha \beta} 
=   
\Phi^{\rm SM}_\alpha \sigma^{\rm SM}_\beta
\sum_{k,l} e^{-i {L \Delta m_{kl}^2 \over 2 E_\nu}}
\nnl 
&&
{\hspace{-3cm}\times}
\left [ U_{\alpha k}^* U_{\alpha l}  
 + p_{XL}(\epsilon_X U)_{\alpha k}^* U_{\alpha l}
 +  p_{XL}^* U_{\alpha k}^* (\epsilon_X U)_{\alpha l}
 + p_{XY} (\epsilon_X U)_{\alpha k}^*(\epsilon_Y U)_{\alpha l}  \right ]
\nnl 
&&
\hspace{-3cm}\times
\left [
U_{\beta k} U_{\beta l}^*  
 +  d_{X'L}(\epsilon_{X'} U)_{\beta k} U_{\beta l}^*
 +  d_{X'L}^* U_{\beta k} (\epsilon_{X'} U)^{*}_{\beta l}
 + d_{X'Y'} (\epsilon_{X'} U)_{\beta k}(\epsilon_{Y'} U)^{*}_{\beta l}  \right], 
 \quad 
\eea
\end{small}
where in this formula repeated $X,Y,X',Y'$ indices are implicitly summed over, and we define the production and detection coefficients as
\beq\label{eq:coefficients}
p_{XY} \equiv   {\int d \Pi_{P'} A_X^P \bar A_Y^P \over \int d \Pi_{P'} |A_L^P|^2 }, 
\quad d_{XY} \equiv  {\int d \Pi_{D} A_X^D \bar A_Y^D \over \int d \Pi_D |A_L^D|^2 }. 
\eeq 
We show in Section~\ref{sec:matching} the expressions of the above coefficients for different  processes. 
For antineutrinos \eref{rategeneraleft} 
holds with $U \leftrightarrow U^*$ and $\epsilon_X \leftrightarrow \epsilon_X^*$. 
The formulas for the neutrino oscillation probability are collected in Section~\ref{sec:oscprob}.  

At  {\em linear level} in $\epsilon$ the QFT expression in \eref{rategeneraleft} matches the QM-NSI one in \eref{ratensi} provided  NSI parameters are 
expressed by the EFT parameters as
\bea
\label{eq:matching}
\epsilon^s_{\alpha \beta} &= &\sum_X p_{XL} [\epsilon_X]^*_{\alpha \beta},
\quad
\epsilon^d_{ \beta\alpha} = \sum_X d_{XL}[\epsilon_X]_{\alpha \beta}.
\eea
Therefore, the QM-NSI formalism can approximate the correct oscillation probability obtained from the general EFT as long as the deviation from the SM, encoded in the coefficients  $[\epsilon_X]_{\alpha \beta}$,  is sufficiently small. 
If non-SM-like interactions are involved (that is with a different Lorentz structure than $V$-$A$), the NSI parameters obtained via the matching in~\eref{matching} may be a function of the neutrino energy and they do not satisfy anymore the relation $\epsilon^s = \epsilon^{d \, \dagger}$ valid for the $V$-$A$ case.

Beyond the linear approximation the QM-NSI formalism fails in general because no matching can be found to connect with the QFT result. 
The consistency condition for the matching in \eref{matching} to be valid to all orders in $\epsilon$ is
\beq
\label{eq:condition}
p_{XL}p_{YL}^*= p_{XY},
\quad
 d_{XL} d_{YL}^*=d_{XY},
\eeq 
for all $X$ and $Y$ for which $\epsilon_{X,Y}$ are non-zero. 
\eref{condition} is trivially satisfied if the only NP deformations are of the $V$-$A$ type, that is if only $\epsilon_L$ is non-zero, in agreement with our previous discussion. 
However, for non-SM-like deformations~\eref{condition} is typically not satisfied, because then
$A_X^{P,D}$ may have different dependence on kinematic variables than $A_L^{P,D}$. 

In the next section we look at specific processes and give concrete examples where \eref{condition} are {\em not} satisfied. We also show cases where the conditions do hold.

\section{Application to specific processes}
\label{sec:matching}

The matching between the NSI parameters $\epsilon^s$ and $\epsilon^d$ to the EFT Wilson Coefficients $\epsilon_X$ depends on the specific processes in which neutrinos are produced or detected, as shown in Eq.~(\ref{eq:matching}).
The process dependence is encoded in the production and detection coefficients 
 $p_{XY}$ and $d_{XY}$ defined in Eq.~(\ref{eq:coefficients}).
With the production and detection coefficients at hand, we can verify whether the consistency condition in \eref{condition} is satisfied. If it is not, the matching is only valid at the linear order in $\epsilon_X$, whereas at higher orders it fails.
In the latter case, the QM-NSI approach does not reproduce the correct dependence on  EFT parameters beyond the linear order in $\epsilon_X$. 
Here we list the production and detection coefficients for inverse beta decay, nuclear decay, and pion decay, and discuss the validity of the matching in each case. The discussion of muon decay is left for Section~\ref{sec:Muondecay}. 
Table~\ref{tab:matching} summarizes the linear matching between the NSI parameters and EFT Wilson coefficients for these processes.

Neutrino detection through non-elastic processes (quasi-elastic, deep-inelastic, or resonances) are more complicated, and even in the SM it is often challenging to provide accurate predictions, mainly due to nontrivial hadronic/nuclear physics involved. For this reason, we leave these processes for future work.

\begin{table}[t]
\centering
\begin{tabular}{|p{6cm}|p{9.4cm}|}
\hline
\bf ~~ Neutrino Process & \bf ~~NSI Matching with EFT
 \\\hline\hline
 \vspace{-0.25cm}&\\
$~\nu_e$ produced in beta decay & 
$~~\epsilon^s_{e \beta}=[\epsilon_L-\epsilon_R-\frac{g_T}{g_A}\frac{m_e}{f_T(E_\nu)}\epsilon_T]^*_{e \beta}$\\
\vspace{-0.25cm}&\\
\hline
\vspace{-0.25cm}&\\
$~\nu_e$ detected in inverse beta decay &
$~~\epsilon^d_{ \beta e}=\Big[\epsilon_L+\frac{1-3g_A^2}{1+3g_A^2}\epsilon_R-\frac{m_e}{E_\nu-\Delta}\big(\frac{g_S}{1+3g_A^2}\epsilon_S-\frac{3g_Ag_T}{1+3g_A^2}\epsilon_T\big)\Big]_{e \beta}$\\
&\vspace{-0.25cm}\\
\hline
\vspace{-0.25cm}&\\
$~\nu_\mu$ produced in pion decay &  
$~~\epsilon^s_{\mu \beta}=[\epsilon_L-\epsilon_R
-\frac{m_\pi^2}{m_\mu(m_u+m_d)}\epsilon_P]^*_{\mu \beta}$\vspace{-0.25cm}\\
&\\
\hline
\vspace{-0.25cm}&\\
$~\nu_\mu$ and $\bar \nu_e$ produced in muon decay \hspace{-2cm}&  
$~~\epsilon^s_{\mu \beta} =  \left [\rho_L  - {3 m_e \over 3 m_\mu - 4 E_\nu } \rho_R \right ]_{ee \beta \mu}$\\ 
& $~~\epsilon^s_{e \beta} =  \left [\rho_L  -  { m_e \over 2 m_\mu - 4 E_{\bar \nu} } \rho_R \right ]_{e \beta \mu \mu}$ 
\vspace{0.25cm}
 \\
\hline
\end{tabular}
\caption{
Summary of the matching between NSI parameters and EFT Wilson coefficients. See~\sref{matching} and~\sref{Muondecay} for further details about the validity of the QM-NSI approach in each case and for the definition of the Wilson coefficients $\rho$ relevant for muon decay. For antineutrinos matching is the same up to complex conjugation. 
}
\label{tab:matching}
\end{table}

A common detection process of low-energy neutrinos is the \textbf{inverse beta decay},  $\nu\, p \to n\, e$. For this case we find the following detection coefficients:
\bea\label{eq:IBDcoefficients}
d_{LL}&=&1,\quad d_{RL}=\frac{1-3g_A^2}{1+3g_A^2},\quad d_{SL}= d_{SR} = -\frac{g_S}{1+3g_A^2}\frac{m_e}{E_\nu-\Delta},
\nnl
 d_{TL}&=& - d_{TR} = \frac{3g_Ag_T}{1+3g_A^2}\frac{m_e}{E_\nu-\Delta},
 \nnl 
d_{RR}&=&1,\quad d_{SS}=\frac{g_S^2}{1+3g_A^2},\quad  d_{TT}=\frac{3g_T^2}{1+3g_A^2},
\quad
\eea
where $g_A = 1.251(33)$, $g_S= 1.02(10)$, $g_T = 0.987(55)$ are the axial, scalar and tensor nucleon charges \cite{Gonzalez-Alonso:2013ura,Chang:2018uxx,Gupta:2018qil,Aoki:2019cca}, $\Delta=m_n-m_p\sim 1.3$ MeV is the neutron-proton mass difference and $m_e$ is the electron mass. We note that the usual chiral factor $\sim m_\ell/E_\nu$ associated to (pseudo-)scalar and tensor interactions~\cite{Kopp:2007ne,Kopp:2009zza} is of order one in this case.

The NSI detection parameters can thus be related to the EFT parameters as  
\bea\label{eq:IBDmatching}
\epsilon^d_{ \beta e} &= &\sum_X d_{XL}[\epsilon_X]_{ e\beta}
\nnl &=&
[\epsilon_L]_{e \beta}+\frac{1-3g_A^2}{1+3g_A^2}[\epsilon_R]_{e \beta}-\frac{m_e}{E_\nu-\Delta}\Big(\frac{g_S}{1+3g_A^2}[\epsilon_S]_{e \beta}-\frac{3g_Ag_T}{1+3g_A^2}[\epsilon_T]_{e \beta}\Big). 
\eea
The consistency condition in \eref{condition} is  satisfied only for the $V$-$A$ case, and fails if other NSIs are present. For example, consider NP of the $V$+$A$ type affecting the process, that is~$[\epsilon_R]_{e\beta} \neq 0$ for some $\beta$. 
Since $|d_{RL}|^2\neq d_{RR}$, we conclude that the effect of $V$+$A$ interactions in neutrino experiments that involve inverse beta decay {\em cannot} be described by the NSI parameters $\epsilon^{s,d}$ beyond the linear level.

In the presence of scalar and tensor interactions we have again $|d_{SL}|^2 \neq d_{SS}$, $|d_{TL}|^2 \neq d_{TT}$. 
Moreover, in these two cases the left-hand sides depend on the neutrino energy, while the right-hand sides do not.

\textbf{Reactor electron antineutrinos} $\bar\nu_e$ are 
produced via beta decays of nuclear fission products. 
To calculate the corresponding amplitudes we assume that only the Gamow-Teller type decays are important (see Ref.~\cite{Falkowski:2019xoe} for further details). With this assumption the non-zero coefficients are
\bea
p_{LL}&=&-p_{RL}=1,\quad 
p_{TL}= - p_{TR} =  -\frac{g_T}{g_A}\frac{m_e}{f_T(E_\nu)},\nnl
p_{RR}&=&1,\quad p_{TT}=\frac{g_T^2}{g_A^2},\quad  
\eea
which gives the following matching with NSI parameters
\bea\label{eq:nuclearmatchin}
\epsilon^s_{e \beta} &= &\sum_X p_{XL} [\epsilon_X]^*_{e \beta}=[\epsilon_L]^*_{e \beta}-[\epsilon_R]^*_{e \beta}-\frac{g_T}{g_A}\frac{m_e}{f_T(E_\nu)}[\epsilon_T]^*_{e \beta}.
\eea
Here $f_T(E_\nu)$ is a function that depends on the nuclear decays taking place in the reactor, which was calculated using certain approximations in Ref.~\cite{Falkowski:2019xoe}. 
We see that the relation in Eq.~(\ref{eq:condition}) is not satisfied for the tensor case: $\frac{m_e^2}{f^2_T(E_\nu)}\neq 1$, which implies that reactor antineutrino production cannot be fully described by the QM-NSI formalism in the presence of tensor interactions.
Moreover, the energy dependence at the linear level (entering via $p_{TL}$) is not there at the quadratic level (because $p_{TT}$ is a constant), which will be missed if we use Eq.~(\ref{eq:nuclearmatchin}) in~\eref{ratensi}. 
For the left- and right-handed interactions the matching is valid at all orders.

We see in this last example that one should not jump into the conclusion that the QM-NSI formalism {\em always} fails for non-SM-like interactions. 
For instance, this is not the case if $ A_X^{P,D} = c_X^{P,D} A_L^{P,D}$, where $c_X^{P,D}$ is a constant independent of the kinematic variables to be integrated over.  
The latter happens e.g. in the 2-body decay of a spin-zero particle. This includes of course the phenomenologically relevant case of neutrino production through \textbf{pion decays}. 
Thanks to the pseudoscalar nature of the pion, 
the only non-zero hadronic matrix elements for this decay 
are 
$\bra{0} \bar u \gamma^\mu \gamma_5 d \ket{\pi^+}$ 
and $\bra{0} \bar u \gamma_5 d\ket{\pi^+}$.  
As a result the production process is sensitive only to axial ($\epsilon_L$-$\epsilon_R$) and pseudo-scalar ($\epsilon_P$) interactions. 

For pions at rest, the non-zero production coefficients are
\bea
\label{eq:pioncoefficients}
p_{LL}&=&-p_{RL}=1,\quad 
p_{PL}= -p_{PR} = {-}\frac{m_\pi^2}{m_\mu(m_u+m_d)},
\nnl 
p_{RR}&=&1,\quad p_{PP}=\frac{m_\pi^4}{m_\mu^2(m_u+m_d)^2}. \quad
\eea
The NSI production parameters can thus be related to the EFT parameters as 
\bea\label{eq:pionmatching}
\epsilon^s_{\mu \beta} = \sum_X p_{XL} [\epsilon_X]^*_{\mu \beta}=[\epsilon_L]^*_{\mu \beta}-[\epsilon_R]^*_{\mu \beta}
-\frac{m_\pi^2}{m_\mu(m_u+m_d)}[\epsilon_P]^*_{\mu \beta}.
\eea
We see that the consistency condition in \eref{condition} is satisfied for all the interactions involved in pion decay. 
Therefore, neutrino production via pion decays can  be described by  the  QM-NSI  formalism  to  all  orders (with the above matching), even in the presence of non-SM-like interactions.

\section{Oscillation probability}
\label{sec:oscprob}

 So far in this paper the basic quantity we have worked with was the event rate $R_{\alpha \beta}$ in \eref{rateqft}, which is an observable in neutrino experiments. 
This quantity can be decomposed into 
the product of the oscillation probability (\eref{oscqft}) and the no-oscillation result (\eref{ratenoscQFT}):
\beq
\label{eq:pdecompose}
R_{\alpha\beta} = P_{\alpha\beta} \times \Phi_\alpha \sigma_\beta~.
\eeq
In this section we the expressions for the oscillation probability $P_{\alpha\beta}$ in both QM-NSI and QFT frameworks, and we also discuss some relevant features.
From the general QFT viewpoint,  the decomposition in \eref{pdecompose} may seem artificial, as the rate in~\eref{rateqft} is directly observable in neutrino experiments. 
Nonetheless, there are advantages of defining the oscillation probability that go beyond its obvious intuitive qualities.  
First,  for the sake of calculating {\em ratios} of measurements at different distances $L$ for a fixed $E_\nu$, the ratio of probabilities is the same as the ratio of rates.  
Second, in many familiar scenarios the physics contributing to the oscillation probability in~\eref{oscqft} and no-oscillation piece in~\eref{ratenoscQFT} is distinct.  
In the SM, electroweak and hadronic parameters contribute to the flux and cross-section, while neutrino masses and mixing angles contribute to the oscillation probability. 
In the EFT scenario (\ref{eq:EFT_lweft}) at the {\em linear} level only flavor off-diagonal $\epsilon_X$ affect the oscillation probability, whereas flavor diagonal $\epsilon_X$ affect the flux/cross section~\cite{Falkowski:2019xoe}. 
It is important to note however that such a separation does not hold beyond the linear level.

In the QM-NSI approach the oscillation probability is given by
\beq
\label{eq:oscnsi}
P_{\alpha \beta}^{\rm QM} 
=
(N_\alpha^s N_\beta^d)^{-2} \sum_{k,l}  
 e^{-i {L \Delta m_{kl}^2 \over 2 E_\nu}}
  [x_s]_{\alpha k} 
  [x_s]^*_{\alpha l}
  [x_d]_{\beta k} 
  [x_d]^*_{\beta l}~.
\eeq
As discussed in~\sref{Formalism}, we have $x_s \equiv (1 + \epsilon^s)U^* $ and $x_d \equiv (1 + \epsilon^d)^T U$, and the normalization factors are 
\beq 
(N_\alpha^s N_{\beta}^d)^2 = 
\left[  (1 +\epsilon^{s})(1+\epsilon^{s\, \dagger}) \right ]_{\alpha\alpha}
\left [  (1 +\epsilon^{d \, \dagger})(1+\epsilon^{d}) \right]_{\beta\beta}.  
\eeq 

In the QFT approach the oscillation probability depends on the parameters of the EFT Lagrangian in \eref{EFT_lweft} as
\begin{small}
\bea
\label{eq:oscgeneraleft}
P_{\alpha \beta}^{\rm  QFT} 
&=&   
N_{\alpha \beta}^{-1}
\sum_{k,l} e^{-i {L \Delta m_{kl}^2 \over 2 E_\nu}}
\nnl &{\times}&
\left [ U_{\alpha k}^* U_{\alpha l}  
 + {\sum_{X}} p_{XL}(\epsilon_X U)_{\alpha k}^* U_{\alpha l}
 + {\sum_{X}} p_{XL}^* U_{\alpha k}^* (\epsilon_X U)_{\alpha l}
 + {\sum_{X,Y}} p_{XY} (\epsilon_X U)_{\alpha k}^*(\epsilon_Y U)_{\alpha l}  \right ]
\nnl &\times&
\left [
U_{\beta k} U_{\beta l}^*  
 + {\sum_{X}} d_{XL}(\epsilon_X U)_{\beta k} U_{\beta l}^*
 + {\sum_{X}} d_{XL}^* U_{\beta k} (\epsilon_X U)^{*}_{\beta l}
 +{\sum_{X,Y}} d_{XY} (\epsilon_X U)_{\beta k}(\epsilon_Y U)^{*}_{\beta l}  \right], \qquad 
\eea
\end{small}
where the coefficients 
 $p_{XY}$ and $d_{XY}$
are defined in \eref{coefficients} and the normalization factor is 
\begin{small}
\bea
\label{eq:oscgeneraleftnorm}
N_{\alpha \beta} 
&=&
\left [ {\bf 1}   
 + {\sum_{X}} p_{XL} \epsilon_X^*
 + {\sum_{X}} p_{XL}^* \epsilon_X 
 + {\sum_{X,Y}} p_{XY} \epsilon_Y \epsilon_X^\dagger\right ]_{\alpha \alpha}
\nnl &{\times}&
\left [ {\bf 1}   
 + {\sum_{X}}d_{XL} \epsilon_X
 + {\sum_{X}} d_{XL}^* \epsilon_X^*
 + {\sum_{X,Y}} d_{XY} \epsilon_X \epsilon_Y^\dagger\right ]_{\beta \beta}
\, .
\eea
\end{small}
The QM-NSI and QFT probabilities can be matched as in \eref{matching} only when the conditions  $p_{XY} =  p_{XL} p_{YL}^*$ and  $d_{XY} =  d_{XL} d_{YL}^*$ are satisfied for each $X$, $Y$ for which $\epsilon_{X,Y}$ are non-zero. 
In the case of $V$-$A$ interactions we have 
$p_{LL} = d_{LL} = 1$ and the consistency conditions are automatically satisfied. 
The SM limit corresponds to $\epsilon_X = 0$ in the EFT, 
or $\epsilon^s = \epsilon^d = 0$ in the QM-NSI  approach, 
in which case we recover the familiar expression 
\beq
P_{\alpha \beta}^{\rm SM} 
=   
\sum_{k,l} e^{-i {L \Delta m_{kl}^2 \over 2 E_\nu}}
 U_{\alpha k}^* U_{\alpha l}   U_{\beta k} U_{\beta l}^*  ~.
\eeq

It is well-known that NP can affect the oscillation probability at zero distance, i.e., ${\mbox P_{\alpha\beta} (L\!=\!0)\neq \delta_{\alpha\beta}}$~\cite{Langacker:1988up}. 
We find that:
\bi
    \item There are no such ``zero-distance effects" at linear order in NP. Let us note that in the $\alpha=\beta$ case the rate itself is affected by linear effects in $[\epsilon_X]_{\alpha \alpha}$, but they come from NP modifying the neutrino flux and detection cross-section in \eref{ratenoscQFT}. 
    \item At quadratic order, zero-distance effects do appear in general.
    \item Zero-distance effects vanish at all orders in the $\alpha=\beta$ case with $V$-$A$ interactions, i.e. ${\mbox P_{\alpha\alpha}^{V-A} (L\!=\!0) = 1}$. 
\ei
Our results are therefore relevant for the study of zero-distance effects since they are quadratic and, in the $\alpha=\beta$ case, necessarily non-SM-like.

 \section{Muon decay}
 \label{sec:Muondecay}
Finally, we consider neutrino production from muon decay (at rest).\footnote{For leptonic tau decays the discussion is completely analogous, up to a trivial change of indices.}
For this purpose we need to extend our EFT so as to describe the low-energy neutrino interactions with leptons. 
At the leading order they can be parametrized as
\beq
\label{eq:muon_eft}
\cL \supset 
-{2 \over v^2}  \left [ 
\left ( \delta_{\alpha a} \delta_{\beta  b}  + [\rho_L]_{ a \alpha  \beta b} \right ) 
 (\bar \ell_a \gamma^\rho P_L \nu_\alpha) (\bar \nu_\beta \gamma_\rho P_L \ell_b)
-2   [\rho_R]_{ a \alpha \beta b} (\bar \ell_a P_L \nu_\alpha )   (\bar \nu_\beta P_R \ell_b) 
\right ],  
\eeq 
where $[\rho_X]_{ a \alpha \beta b}^* =[\rho_X]_{ b  \beta \alpha a}$ for the Lagrangian to be Hermitian. 
A complication in muon decay is that both a neutrino and an antineutrino are produced in the same process. 
In the following we present the rate $R$ of detecting a neutrino of flavor $\beta$ summed over all antineutrino eigenstates, and the rate $\bar R$ of detecting an antineutrino of flavor $\beta$ summed over all neutrino eigenstates.   
For simplicity, here we neglect new physics in detection.\footnote{The $R_{\mu\beta}$ expression including NP in detection is simply obtained replacing the SM detection piece $U_{\beta k} U_{\beta l}^*$ by the BSM detection piece shown in the second line of~\eref{rategeneraleft}, and likewise for $\bar{R}_{e\beta}$ with trivial changes.} 
We find 
\begin{small}
\bea 
\label{eq:muon_R}
R_{\mu \beta} 
&=& 
\Phi^{\rm SM}  \sigma^{\rm SM}_\beta
\sum_{k,l} e^{-i {L \Delta m_{kl}^2 \over 2 E_\nu}}
U_{\beta k} U_{\beta l}^*
\nnl &{\times} &  
\bigg [  U_{\mu k}^*   U_{\mu l} 
 + p_{XL}  [\rho_X U^*]_{e e \mu k}   U_{\mu l} 
 + p_{LX} U_{\mu k}^*   [\rho_X U^*]_{ee \mu l}^* 
+\sum_\gamma 
 p_{XY} [\rho_X U^*]_{e \gamma\mu k}    [\rho_Y U^* ]_{e\gamma \mu l}^* 
  \bigg ]
~, 
\nnl
\bar R_{e \beta} 
&=& 
\bar \Phi^{\rm SM}  \bar \sigma^{\rm SM}_\beta
\sum_{k,l} e^{-i {L \Delta m_{kl}^2 \over 2 E_\nu}}
U_{\beta k}^* U_{\beta l}
\nnl &{\times} &  
\bigg [  
 U_{e k}   U_{e l}^*   
 +  \bar  p_{XL} [\rho_X U ]_{e \mu \mu k}  U_{e l}^*  
 + \bar p_{LX}  U_{e k}  [\rho_X U ]_{e \mu \mu l}^*  
 + \sum_\gamma \bar p_{XY}   [\rho_X U ]_{e \gamma \mu k}   [\rho_Y U ]_{e \gamma \mu l}^*
 \bigg ] ~, ~~~~~
\eea
\end{small}
where we have defined the matrix contractions 
$[\rho_X U^*]_{a \alpha b k} \equiv \sum_\beta  [\rho_X]_{a \alpha \beta  b} U^*_{\beta k}$ and 
$[\rho_X U ]_{a \gamma b k}  \equiv \sum_\alpha [\rho_X ]_{a \alpha \gamma b} U_{\alpha k}$. 
The SM cross sections for detecting a neutrino and an antineutrino of flavor $\beta$ are denoted as $\sigma^{\rm SM}_\beta$ and $\bar \sigma^{\rm SM}_\beta$ respectively.
The differential neutrino and antineutrino fluxes in the SM limit are \beq
\label{eq:muon_flux}
\Phi^{\rm SM}  = {N_\mu \over 4\pi L^2} {m_\mu E_\nu^2 (4 E_\nu - 3 m_\mu)  \over 24 \pi^3 v^4} + \cO(m_e),
\qquad 
\bar \Phi^{\rm SM} = {N_\mu \over 4\pi L^2} {m_\mu E_{\bar \nu}^2 (m_\mu - 2 E_{\bar \nu} )  \over 4 \pi^3 v^4} + \cO(m_e),
\eeq 
where $N_\mu$ are the number of muons. 
Note that the two are different functions of the (anti)neutrino energy, which is due to the structure of the muon decay matrix element. The production coefficients are given by 
\beq
p_{RL}  = p_{LR}  = - {3 m_e \over 3 m_\mu - 4 E_\nu } + \cO(m_e^2), \qquad 
p_{RR} =  {6 m_\mu - 12 E_\nu \over 3 m_\mu - 4 E_\nu }+ \cO(m_e). 
\eeq 
\beq 
\bar p_{RL}  = \bar p_{LR} =  - { m_e \over 2 m_\mu - 4 E_{\bar \nu} } , \qquad 
\bar p_{RR} =  {3 m_\mu - 4 E_{\bar \nu} \over 6 m_\mu - 12 E_{\bar \nu} }+ \cO(m_e).  
\eeq 
and $p_{LL} = \bar p_{LL} = 1$.
Note that $p_{RL}$ and $\bar p_{RL}$, 
which control the linear effects proportional to the $\rho_R$ Wilson coefficients, are suppressed by $m_e/m_\mu \approx 0.005$.
This makes the quadratic terms dominant unless $\rho_R$ are strongly suppressed, below $\cO(1\%)$, in which case they are probably too small to be observable in current experiments anyway. 
This fact amplifies the need for a correct treatment of  quadratic terms, as ensured by our QFT formalism.   

To discuss whether the above-given QFT result can be matched to the QM-NSI formalism one needs to specify how the latter describes (anti)neutrino production from muon decay~\cite{Antusch:2006vwa,Tang:2017qen,C.:2019dbf}. 
Works carried out within the ``non-unitary mixing matrix" setup~\cite{Antusch:2006vwa} introduce in~\eref{ratensi} an additional normalization factor, $(N^s_\delta)^2$, associated to the (anti)neutrino $\nu_\delta$ that is emitted and not detected. In the $R_{\mu\beta}$ case discussed above, \eref{ratensi} would be replaced by
\bea
\label{eq:ratensiMuon}
R_{\mu \beta}^{\rm QM}
=
\Phi^{\rm SM}_\mu \sigma^{\rm SM}_\beta 
P_{\mu \beta}^{\rm QM} 
(N_e^s N_\mu^s N_\beta^d)^2
~.
\eea 
The expression for the QM oscillation probability is not changed, and thus the additional normalization factor, $(N^s_e)^2$, does \emph{not} cancel in the observable rate.
One can check that the QFT rate in~\eref{muon_R} does reproduce such result at all orders only if the effective interactions are $V$-$A$ \emph{and} can be factorized as follows
\bea
\label{eq:factorization}
\delta_{\alpha a} \delta_{\beta  b}  + [\rho_L]_{ a \alpha  \beta b} = 
\left( \delta_{\alpha a} + \omega^*_{ a \alpha} \right) 
\left( \delta_{\beta b} + \omega_{ b \beta} \right)~.
\eea
In such a case, the matching is given by $\epsilon^s=\omega$. Such factorization holds in particular (but not only) when the BSM low-energy 4-lepton interaction is generated by modifications of the coupling of the $W$ boson to leptons, as e.g. in the ``non-unitary mixing matrix" setup. However, if the $V$-$A$ Wilson coefficient does not satisfy~\eref{factorization}, or if other interactions are present, the QM-NSI prescription in~\eref{ratensiMuon} fails, even at linear order. That is, there is no matching between $\epsilon^s$ and $\rho_X$ such that \eref{ratensiMuon} is recovered from the QFT result in~\eref{muon_R}.

One can formulate the QM-NSI approach in a different form, e.g. using~\eref{ratensi} without any modification
\bea
R_{\mu \beta}^{\rm QM}
= 
\Phi^{\rm SM}_\mu \sigma^{\rm SM}_\beta 
P_{\alpha \beta}^{\rm QM} 
(N_\mu^s N_\beta^d)^2~\nonumber
~.
\eea 
In that case it is possible to find a linear matching valid in general, namely
\beq
\epsilon^s_{\mu \alpha} = \sum_X p_{XL} [\rho_X]_{ee \alpha \mu}, \qquad 
\epsilon^s_{e \alpha} =  \sum_X \bar p_{XL} [\rho_X]_{e \alpha \mu \mu}. 
\eeq 
However, this approach comes at the cost of (i) not being able to describe the relevant case of~\eref{factorization} at all orders; and (ii) having process-dependent coefficients even in the $V$-$A$ case.

All in all, we conclude that (anti)neutrino production through muon decay presents additional features with respect to the production through a semileptonic process. 
As a result, the limitations of the QM-NSI approach are even more severe. In particular we see that a generic $V$-$A$ interaction cannot be described exactly through the QM-NSI formalism.

In closing we note that the relation between  the  parameter $v$ and the Fermi constant $G_F$ becomes more complicated in the presence of the effective interactions in \eref{muon_eft}:
\beq
G_F^2  = {1 \over 2 v^4}   \left [1 +2  \re [\rho_L]_{e e \mu \mu}  + \sum_{\alpha \beta} \left ( [\rho_L]_{e \alpha \beta \mu} [\rho_L]_{e \alpha \beta \mu}^*    + [\rho_R]_{e \alpha \beta \mu}  [\rho_R]_{e \alpha \beta \mu}^* \right )  \right ] + \cO(m_e) . 
\eeq  
This is because the same Wilson coefficients that affect oscillations of neutrinos from muon decay also affect the muon lifetime, from which $G_F$ is determined experimentally. 
This has to be taken into account in order to derive consistent constraints on $\rho_X$ if absolute rates are used in the oscillation analysis. 
If on the other hand only ratios of rates at different distance $L$ are used, then $v$ cancels out in the observables, and this subtlety is irrelevant.


\section{ Discussions and Conclusions}
\label{sec:conclusion}
We close with several
comments on the results derived above:

{\bf 1.} In this paper we only discussed charged-current NSI and assumed the absence of matter effects in propagation.  
The neutral-current NSI {\em other} than the matter effects can also be correctly described by QFT expressions analogous to Eqs.~(\ref{eq:rateqft})-(\ref{eq:oscqft}), 
and they are relevant e.g. if neutrinos are detected via coherent scattering on nuclei. 
To include NSI entering via the matter effects one would need to modify the neutrino propagator in the derivation in Appendix~\ref{app:oscqft} starting from Eq.~(\ref{eq:NOF_neutrinopropagator}).  
We leave this for future work. 

{\bf 2.}  It is worth stressing that charged-current NSI  modify not only  the flux and cross-section in~\eref{ratenoscQFT}, but also the oscillation probability in vacuum. 
The latter follows directly from \eref{oscqft}, which depends on the production and detection amplitudes. 
Generically, that dependence does {\em not} cancel between the numerator and denominator in \eref{oscqft}. 

{\bf 3.} 
 The observable in \eref{rateqft} may depend on NP in two distinct ways. 
 One is {\em direct}, e.g. through a dependence of the production and detection amplitudes ${\cal M}^{P,D}_{\alpha k}$ on the NP parameters $\epsilon_X$ of the Lagrangian in \eref{EFT_lweft}.  
 The other is {\em indirect}, due to  NP ``polluting'' the observable used for determination of the SM parameters~\cite{Descotes-Genon:2018foz}. 
 This is the case for the CKM parameter $V_{ud}$ in \eref{EFT_lweft}. If NP is present, $\beta$ decay experiments determine a combination of $V_{ud}$ and $[\epsilon_X]_{e\beta}$ parameters,  and in this case the value of $V_{ud}$ cannot be just taken from PDG.  
 This indirect effect is ignored in most of the prior neutrino literature, even though it is of the same order as the direct effects, leading to incorrect results. 
For instance, indirect and direct effects generated by the coefficient $[\epsilon_L]_{ee}$ cancel at all orders, making this coefficient unobservable in oscillation experiments~\cite{Falkowski:2019xoe}.

The main results of this paper are: i) The expression in~\eref{rategeneraleft} for the event rate in neutrino oscillation experiments including nonstandard  charged-current interactions described by the EFT  Lagrangian in~\eref{EFT_lweft};
ii) The matching in~\eref{matching} and~\tref{matching}, valid at the linear level in NP, between the EFT coefficients that describe the underlying interactions and the QM-NSI parameters; 
iii) The consistency condition in~\eref{condition} for that matching (and the simplified QM-NSI approach itself) to be valid to all orders in NP parameters.

Our results are particularly relevant for analyses of oscillation data when effects of non-SM-like physics (or equivalently $\epsilon^s \neq \epsilon^{d\,\dagger}$) are taken into account~\cite{Kopp:2007ne,Ohlsson:2008gx,Biggio:2009nt,Leitner:2011aa,Ohlsson:2012kf,Agarwalla:2014bsa,Li:2014mlo,Blennow:2015nxa}. 
We give a more fundamental meaning to the long list of existing analyses of oscillation data carried out within the traditional QM-NSI approach. 
Their discovery potential can now be consistently analyzed and compared, 
among themselves and together  with non-oscillation probes that are sensitive to the same underlying physics. Even at the linear level in $\epsilon_X$ we do find important and measurable effects that are not captured by the standard NSI formalism. 
Namely, in the presence of non-SM-like interactions in the EFT Lagrangian, we find that the NSI parameters depend on the neutrino energy.
This dependence, which has several phenomenological consequences  has never been considered before within the standard NSI formalism.


\begin{acknowledgments}

We thank Joachim Kopp for valuable discussions.
We also acknowledge the hospitality of LPT Orsay, CERN-TH, and Karlsruhe Institute of Technology, where this work was developed.
MGA is supported by the {\it Generalitat Valenciana} (Spain) through the {\it plan GenT} program (CIDEGENT/2018/014).
ZT is supported by Funda\c{c}\~{a}o de Amparo \`{a} Pesquisa do Estado de S\~{a}o Paulo (FAPESP) under contract 2018/21745-8.
AF is partially supported by the European Union's Horizon 2020 research and innovation programme under the Marie Sk\l{}odowska-Curie grant agreements No 690575 and No 674896.

\end{acknowledgments}

\appendix

\section{Oscillations in QFT}
\label{app:oscqft}

In this appendix we derive the master formula in \eref{rateqft} describing the number of neutrino events detected at a distance $L$ from the source, taking into account possible neutrino oscillations and nonstandard charged-current interactions.   
Our approach follows similar steps as Ref.~\cite{Giunti:1993se}, but with some notable modifications. 
The  main differences are: 1) we allow for general non-SM charged-current interactions in neutrino production and detection;  2) we work with time-independent\footnote{%
This assumption greatly simplifies the derivation below. 
However, it will lead to a subtlety at one point, since the source is necessarily unstable.}
wave packets for the source and target particles;   
3) we do not assume any particular (e.g. Gaussian) shape of the wave packet.

We consider an experimental setup where a particle $\nu$ is produced in the process $S \to X_\alpha \nu$, and detected via the process $\nu T \to Y_\beta$.  
Here  $X_\alpha$ and $Y_\beta$ are $n_x$- and $n_y$-body final states  ($n_i \geq 1$). 
The indices $\alpha$ and $\beta$  indicate that these states contain charged lepton $\ell_\alpha$ and $\ell_\beta$ respectively, but otherwise their precise identity is irrelevant for this discussion. 
$S$  and $T$  are both one-body particle states  localized away from each other in the coordinate space, describing the neutrino source (e.g. a beta-decaying nucleus in a reactor) and target (e.g. a proton in a detector). 
The particle $\nu$ can be anything: a neutrino, active or sterile, an antineutrino, an axion, an axino, etc., but for simplicity in this section we always refer to it simply  as a {\em neutrino}. 

The idea is to treat the neutrino production and detection together as a single process~\cite{Giunti:1993se}:
\beq
\label{eq:QFTALT_eq1}
S T \to X_\alpha Y_\beta,  
\eeq 
rather than consider the neutrino production and detection separately.
In this approach, neutrino is merely an intermediate particle in the amplitude. 
We will work in the time-independent approximation where the states $S$  and $T$ are represented by wave-packets of  scattering in-states which do not change in time:
\beq
\label{eq:QFTALT_eq2}
 |S T \rangle    =    \int d\Phi(p_S) d\Phi(p_T)
 f_S (p_S) f_T (p_T)  e^{i b p_T}| \mathbf{p}_S \mathbf{p}_T \rangle_{\rm in}, 
\eeq
where $d\Phi(p_j) \equiv {d^3 p_j \over (2\pi)^3 2 E_j}$ for $j=S,T$, 
and $E_j= \sqrt{m_j^2 + |\mathbf{p}_j|^2}$. 
Here, $|\mathbf{p}_j \rangle_{\rm in}$ are momentum eigenstates normalized as   $\langle \mathbf{q}_j  |\mathbf{p}_j\rangle = (2 \pi)^3 2 E_j \delta^3(\mathbf{p}_j - \mathbf{q}_j)$. 
The functions $f_j(p_j)$ are the wave packets describing quantum uncertainty of the momenta of the source and target particles.  
The momentum spread of the wave packets is denoted as  $1/\sigma$, and we assume $\sigma \gg 1/E_{S,T}$.  
We use the normalization 
$\int d\Phi(p) |f_j(p)|^2  = 1$,  which implies  $\langle S T |S T \rangle = 1$. 
The spatial separation between the source and the target is described by the 4-vector $b^\mu$, which is space-like, $b^2 <0$, and plays a similar role as the impact parameter in classical scattering. 
The outgoing states are approximated by pure momentum eigenstates with the eigenvalues $\mathbf{k}_i$, 
$i = 1 \dots n$, where $n = n_x + n_y$ is the number of particles in the final state. 

We are interested in the transition probability for the 
$S T \to X_\alpha Y_\beta$ process: 
\beq
\label{eq:QFTALT_eq3}
N_{\alpha \beta}   = \int\Pi_{i=1}^n d \Phi(k_i)   \left | {}_{\rm out}\langle k_1 k_2 \dots k_n| S T \rangle \right |^2.
\eeq   
Plugging the wave packets for the initial states, and using ${}_{\rm out}\langle k_1 k_2 \dots k_n| p_S p_T \rangle_{\rm in} = (2\pi)^4 \delta^4(p_S +p_T - \sum k_i) \cM$ we obtain 
\bea
\label{eq:QFTALT_eq4}
N_{\alpha \beta}  &=&   \int 
d\Phi(p_S) d\Phi(p_T) d\Phi(p_S') d\Phi(p_T')
 f_S (p_S) f_T (p_T)  f_S (p_S')^* f_T (p_T')^*  e^{i b (p_T-p_T')}
\nnl
 &&{\times} 
(2\pi)^4\delta^4(p_S' + p_T'- p_S - p_T)  
 d \Pi_n  \cM \bar \cM', 
\eea 
where $d \Pi_n \equiv  d \Pi_n (p_S+p_T|k_1 \dots k_n) =  (2 \pi)^4  \delta^4(p_S + p_T - \sum k_i) \Pi_i d \Phi(k_i) $  is the $n$-body phase space for the final-state particles, and $\cM \equiv \cM(p_S p_T \to k_1 \dots k_n)$, $\cM' \equiv \cM(p_S' p_T' \to k_1 \dots k_n)$ are the usual amplitudes calculated by Feynman diagrams. 
Tacitly, $N_{\alpha \beta}$ involves sum/average over all non-observed degrees of freedom, such as polarizations of the initial- and final-state particles. 
It is convenient to change variables as 
$\mathbf{p_S'} = \mathbf{p_S} + \mathbf{r}$, 
$\mathbf{p_T'} = \mathbf{p_T} + \mathbf{r'}$, 
and use $\delta^3(\mathbf{r}+\mathbf{r'})$ to integrate over $d^3 r'$: 
\bea
\label{eq:QFTALT_eq5}
N_{\alpha \beta}  &=&  {1 \over 16 \pi^2} \int  d \Pi_n 
d\Phi(p_S) d\Phi(p_T) d^3 r 
 f_S (p_S) f_T (p_T)  f_S (p_S+r)^* f_T (p_T-r)^*  e^{i b r}
\nnl
 &&{\times} 
{\delta(E_S' + E_T'- E_S - E_T)  \over E_S' E_T'}  \cM \bar \cM', 
\eea 
where $E_S' = \sqrt{(\mathbf{p_S+r})^2 + m_S^2}$, 
$E_T' = \sqrt{(\mathbf{p_T-r})^2 + m_T^2}$. 
The ``momentum mismatch" $\mathbf{r}$, as it is called in \cite{Kosower:2018adc}, 
parametrizes the quantum jitter that will ultimately make oscillations possible. 
We also define $r_0 = E_T - E_T' = E_S' - E_S$. 

Up to this point, our steps have been similar to the classic derivation of the cross section formula, see e.g. Ref.~\cite{Peskin:1995ev}.  
What distinguishes the case at hand is the particular choice of the initial state $\ket{ST}$ describing two spatially separated particles (rather than head-on beams as in the cross section case). 
Furthermore, the amplitude is dominated by the kinematic region(s) where the intermediate $\nu$  is close to the mass shell.
In that region we can represent the amplitude as 
\beq
\label{eq:NOF_neutrinopropagator}
\cM  = \sum_k { \cM(p_S \to k_{X_\alpha} q_{\nu_k})  \cM(q_{\nu_k} p_T \to k_{Y_\beta}) \over q^2 - m_k^2 + i \epsilon} 
+ \cM_{\rm reg}  +  \cM_{\rm loop} , 
\eeq 
where $q \equiv p_S - k_X = k_Y  - p_T$.
The index $k$ counts all active and sterile neutrino and antineutrino states that can be exchanged in this process\footnote{%
In a lepton number conserving theory the exchanged particle is {\em either} a neutrino {\em or} an antineutrino,  however both can contribute at the same time in an $L$-violating theory.} 
and $m_k$ denotes the masses of these states.
If the neutrinos are Dirac the sum goes also over the helicities.
The production and detection amplitudes, 
$\cM_{\alpha k}^P \equiv  \cM(p_S \to k_{X_\alpha} q_{\nu_k}) $ and  $\cM_{\beta k}^D \equiv   \cM(q_{\nu_k} p_T \to k_{Y_\beta}) $ are evaluated for all particles on-shell, including $\nu_k$. 
Such factorization of the residue of the $q^2 = m_k^2$ pole into  production and detection parts is required by unitarity and locality. 
$\cM_{\rm reg}$ denotes the part of the tree-level amplitude that is regular in the $q^2 \to m_k^2$ limit, 
while $\cM_{\rm loop}$ is the part suppressed by powers of $1/16\pi$.    
In the following we will neglect $\cM_{\rm reg}$ and $\cM_{\rm loop}$. 
Inserting the first term of  \eref{NOF_neutrinopropagator}  into \eref{QFTALT_eq5} we get 
\bea
\label{eq:QFTALT_eq7}
N_{\alpha \beta}  &=&  {1 \over 32 \pi^3} \int   d \Pi_n d\Phi(p_S) d\Phi(p_T) d^3 r 
 f_S (p_S) f_T (p_T)  f_S (p_S+r)^* f_T (p_T-r)^*  e^{i b r}  
\nnl
 &\times & 
 {\delta(E_S' + E_T'- E_S - E_T)  \over E_S' E_T'}   \sum_{kl} 
 {\cM_{\alpha k}^P \cM_{\beta k}^D   \over q^2 - m_k^2 + i \epsilon} 
  {\bar \cM_{\alpha l}^{'P} \bar \cM_{\beta l}^{'D}  \over (q + r)^2 - m_l^2 - i \epsilon} . 
\eea 
Trading the energy delta function for a time integral leads to
\bea
\label{eq:QFTALT_eq7b}
N_{\alpha \beta}  &=&  {1 \over 32 \pi^3} \int  dt   d \Pi_n d\Phi(p_S) d\Phi(p_T) d^3 r 
 f_S (p_S) f_T (p_T)  f_S (p_S+r)^* f_T (p_T-r)^*  e^{i b r}
\nnl
 &\times & 
{ e^{i (E_S' + E_T' - E_S- E_T) t} \over E_S' E_T'} 
  \sum_{kl} 
 {\cM_{\alpha k}^P \cM_{\beta k}^D   \over q^2 - m_k^2 + i \epsilon} 
  {\bar \cM_{\alpha l}^{'P} \bar \cM_{\beta l}^{'D}  \over (q + r)^2 - m_l^2 - i \epsilon} . 
\eea 
Instead of integrating the result over time,\footnote{%
If we continued with the time integral, we would find a singular result. 
This  could be expected: since we use the approximation of time-independent wave packets, the rate for the 
$S T \to X_\alpha Y_\beta$ process integrated from $t=-\infty$ to $t = + \infty$ is infinite.
In a physical situation, however, $S$ is unstable, appears at a finite time $t_0$, and decays after a finite time $t_0 + T$. 
Outside this window the process $S T \to X_\alpha Y_\beta$  cannot occur.} 
we ``divide" both sides of the above equation by $dt$. 
In other words,  we are interested in  calculating the {\em event rate}, that is the number of events per unit time: 
\bea
\label{eq:QFTALT_eq7.5}
{d N_{\alpha \beta} \over d t}  &=&  {1 \over 32 \pi^3} \int  d \Pi_n d\Phi(p_S) d\Phi(p_T) d^3 r 
 f_S (p_S) f_T (p_T)  f_S (p_S+r)^* f_T (p_T-r)^*  e^{i b r}
\nnl
 &\times & 
{ e^{i (E_S' + E_T' - E_S- E_T) t} \over E_S' E_T'} 
  \sum_{kl} 
 {\cM_{\alpha k}^P \cM_{\beta k}^D   \over q^2 - m_k^2 + i \epsilon} 
  {\bar \cM_{\alpha l}^{'P} \bar \cM_{\beta l}^{'D}  \over (q + r)^2 - m_l^2 - i \epsilon} . 
\eea 
We now use the formula
\beq
\label{eq:QFTALT_pvFormula}
{1 \over x  - a + i \epsilon} {1 \over x  -  b  -  i \epsilon} 
= - {i \pi \delta(x-a) \over  x  -  b  -  i \epsilon} +  {i \pi \delta(x-b) \over  x - a  +  i \epsilon}
+  {\cal P}{\rm V} \bigg (  {1 \over (x  - a)(x  -  b)}  \bigg ) , 
 \eeq 
where $ {\cal P}{\rm V}$ stands for the principal value.   
This is a simple generalization of the textbook formula 
 ${1 \over x  - a + i \epsilon}   = -i \pi \delta(x-a)  +  {\cal P}{\rm V} \bigg (  {1 \over x  - a}\bigg )$.  
The proof of \eref{QFTALT_pvFormula} is analogous: 
one divides the integration range into 3 sections 
$[-\infty, a-\epsilon]$, $[a+\epsilon,b-\epsilon]$, $[b+\epsilon, \infty]$ (from which the ${\cal P}{\rm V}$ part arises),
and two small half-circles going around the poles  (which yield the delta functions).   
In the following we ignore the ${\cal P}{\rm V}$ part, which is equivalent to keeping only the contribution from on-shell neutrinos. 
The rate will thus be the sum of the contributions from the
 $q^2 = m_k^2$ and  $(q+r)^2 = m_l^2$ poles: 
${d N_{\alpha \beta} \over d t}  = R_{\alpha \beta}^{(1)} + R_{\alpha \beta}^{(2)}$,  
where
\bea
\label{eq:QFTALT_eq9}
R_{\alpha \beta}^{(1)}   &=&  -{i \over 32 \pi^2} \int    d \Pi_n d\Phi(p_S) d\Phi(p_T) d^3 r 
 f_S (p_S) f_T (p_T)  f_S (p_S+r)^* f_T (p_T-r)^*  e^{i b r}
\\
 &\times & 
{ e^{i (E_S' + E_T' - E_S- E_T) t} \over E_S' E_T'}
  \sum_{kl}  { \delta(q^2 - m_k^2)  \big (  \theta(q^0) +   \theta(-q^0) \big )    \over (q + r)^2 - m_l^2 - i \epsilon}  
\cM_{\alpha k}^P \cM_{\beta k}^D  \bar \cM_{\alpha l}^{'P} \bar \cM_{\beta l}^{'D}   , 
  \nnl 
R_{\alpha \beta}^{(2)}   &=&  {i \over 32 \pi^2} \int    d \Pi_n d\Phi(p_S) d\Phi(p_T) d^3 r 
 f_S (p_S) f_T (p_T)  f_S (p_S+r)^* f_T (p_T-r)^*  e^{i b r}
\nnl
 &\times & 
{ e^{i (E_S' + E_T' - E_S- E_T) t} \over E_S' E_T'}
  \sum_{kl}  { \delta((q\!+\!r)^2 - m_l^2)  \big (  \theta(q^0\!+\!r^0) +   \theta(-q^0 \!-\!r^0) \big )      \over q^2 - m_k^2 + i \epsilon}  
 \cM_{\alpha k}^P \cM_{\beta k}^D   \bar \cM_{\alpha l}^{'P} \bar \cM_{\beta l}^{'D}  .
\nonumber
\eea
In fact,  $R_{\alpha \beta}^{(2)}$ is the complex conjugate of $R_{\alpha \beta}^{(1)}$.
To show this, in  $R_{\alpha \beta}^{(2)}$ one shifts the integration variables, 
$\mathbf{p_S} \to \mathbf{p_S}- \mathbf{r}$, 
$\mathbf{p_T} \to \mathbf{p_T} + \mathbf{r}$, 
and then flips $\mathbf{r} \to -\mathbf{r}$. 
Furthermore, in $R_{\alpha \beta}^{(1)}$ in  \eref{QFTALT_eq9} we used 
$1 = \theta(q^0) +   \theta(-q^0)$.  
For $q_0 < 0$ we have  $p_S^0 <  k_X^0$ and $p_T^0 > k_Y^0$. 
This corresponds to the mirror process where the {\em target} decays emitting an antiparticle of $\nu$, which then hits the source. 
Since we are interested in  the source decaying into a neutrino that hits the target, 
we can drop  $\theta(-q_0)$,  and   keep only the $\theta(q_0)$ part. 
For the same reason we drop  $\theta(-q^0 -r^0)$ in  $R_{\alpha \beta}^{(2)}$.
All in all, 
\bea
\label{eq:QFTALT_eq9b}
{d N_{\alpha \beta} \over d t}   &=&  -{i \over 32 \pi^2} \int    d \Pi_n d\Phi(p_S) d\Phi(p_T) d^3 r 
 f_S (p_S) f_T (p_T)  f_S (p_S+r)^* f_T (p_T-r)^*  e^{i b r}
\nnl
 &\times & 
{ e^{i (E_S' + E_T' - E_S- E_T) t} \over E_S' E_T'}
  \sum_{kl}  { \delta(q^2 - m_k^2)   \theta(q^0)   \over (q + r)^2 - m_l^2 - i \epsilon}  
\cM_{\alpha k}^P \cM_{\beta k}^D   \bar \cM_{\alpha l}^{'P} \bar \cM_{\beta l}^{'D}  + \hc . 
\eea 
Note that the expression for the rate is explicitly real, as it should. 
We now use the following identity for the phase space:
\bea
\label{eq:QFTALT_eq10}
 & 
2 \pi \delta((p_1 - k_1 \dots - k_{n_x})^2 - m^2)   \theta(p_1^0 - k_1^0 \dots - k_{n_x}^0) d \Pi_n (p_1+p_2| k_1 \dots k_n)  
& \nnl & 
=   d \Pi_{n_x+1} (p_1| k_1 \dots k_{n_x} q) d\Pi_{n-n_x} (p_2 +q| k_{n_x+1} \dots k_n) . 
\eea 
To prove it, one can rewrite the right-hand side as
\bea & &
d \Pi_{{n_x}+1} (p_1| k_1 \dots k_{n_x} q) d\Pi_{n-{n_x}} (p_2 +q| k_{{n_x}+1} \dots k_n) 
 \nnl  &&
= (2 \pi)^8 \delta^4(p_1 - k_1 \dots -k_{n_x} - q)  \delta^4(p_2 +q  - k_{{n_x}+1}\dots -k_n)     {d^4 q \over (2 \pi)^3} \delta(q^2 -m^2) \theta(q_0)  d \Phi(k_1) \dots d \Phi(k_n)
 \nnl  &&
= (2 \pi)^5  \delta^4(p_1 + p_2 - k_1 \dots -k_n )  \delta^4(p_1 - k_1 \dots -k_{n_x} - q)  d^4 q  \delta(q^2 -m^2) \theta(q_0)  d \Phi(k_1) \dots d \Phi(k_n)
 \nnl  &&
= (2\pi)^5 \delta^4(p_1 + p_2 - k_1 \dots -k_n )  d \Phi(k_1) \dots d \Phi(k_n)  \delta(q^2 -m^2) \theta(q_0)|_{q = p_1 - k_1 \dots -k_{n_x}} . 
\eea 
Applying  \eref{QFTALT_eq10} in  \eref{QFTALT_eq9b} we get
\bea
{d N_{\alpha \beta}  \over dt}  &=&  {1 \over 64 \pi^3} \sum_{kl} \int   d \Pi_P d \Pi_D 
d\Phi(p_S) d\Phi(p_T) d^3 r 
 f_S (p_S) f_T (p_T)  f_S (p_S+r)^* f_T (p_T-r)^* 
\\
 &\times & \cM_{\alpha k}^P \cM_{\beta k}^D 
 \bar \cM_{\alpha l}^{'P} \bar \cM_{\beta l}^{'D}
{ e^{i (E_S' + E_T' - E_S- E_T) t} \over E_S' E_T'} 
{ i  e^{i b r} \over  2 \mathbf{q}\mathbf{r} - \Delta m_{kl}^2  - 2 r_0 \sqrt{\mathbf{q}^2+ m_k^2}  - r^2  + i \epsilon}  + \hc, 
\nonumber
\eea 
 where $\Delta m_{kl}^2 = m_k^2 - m_l^2$.  
Above we abbreviate the production and detection phase spaces as
 $d \Pi_P  \equiv  d \Pi_{{n_x}+1} (p_S| k_1 \dots k_{n_x} q)$,  
  $d \Pi_D \equiv  d\Pi_{n-{n_x}} (p_T +q| k_{{n_x}+1} \dots k_n)$, and $q$ is the 4-momentum of the neutrino. We can simplify this formula by getting rid of the wave packets. 
To this end we assume that the $d^3 r$ integral is dominated by the region $|\mathbf{r}| \ll 1/\sigma$. 
Moreover we assume that the wave packets are sufficiently wide (in momentum space) so that they overlap near the pole $|\mathbf{r}| \sim \Delta m_{kl}^2/|\mathbf{q}|$, i.e., $\sigma \Delta m_{kl}^2 \ll |\mathbf{q}|$. 
If that is the case, we can approximate 
$f_j (p+r) \approx f_j(p)$. 
Furthermore, 
we assume that the wave packets are sharply peaked at classical values  $\bar p_S$ and $\bar p_T$ of the initial momenta, 
in which case we can approximate 
$ |f_j (p_j)|^2 \approx (2 \pi)^3 2 E_j \delta^3(\mathbf{p}_j - \bar p_j)$ and integrate over $d^3 p_S$ and $d^3 p_T$. 
We obtain\footnote{%
Alternatively, we could choose a specific wave packet with the momentum spread of order $1/\sigma$, for example a Gaussian one $f \sim \exp \big (\sigma^2 (\mathbf{p} - \bar p)^2 \big )$.  
This would lead to the same result for the rate up to calculable $\cO\big(\big (\sigma\Delta m_{kl}^2/E_\nu \big )^2\big)$ corrections.  
}
\bea
\label{eq:NOF_tmp1}
{d N_{\alpha \beta} \over dt }  &=&  {1 \over  64 \pi^3} \sum_{kl} \int   d \Pi_P d \Pi_D  d^3 r   \cM_{\alpha k}^P \cM_{\beta k}^D 
 \bar \cM_{\alpha l}^{'P} \bar \cM_{\beta l}^{'D}
\\
 &\times &
{ e^{i (E_S' + E_T' - E_S- E_T) t} \over E_S' E_T'} 
{ i  e^{i b r} \over  2 \mathbf{q}\mathbf{r} - \Delta m_{kl}^2  - 2 (E_T - E_T') \sqrt{\mathbf{q}^2+ m_k^2}  - r^2  + i \epsilon}  + \hc .
\nonumber
\eea

At this point we introduce a number of approximations that are appropriate for the description of broad classes of real-life neutrino experiments: 
\ben
\item  We work in the frame where the target is at rest: 
$\bar p_T = 0$, $E_T = m_T$. 
\item The source and the target are separated by a macroscopic distance $L$. 
For concreteness, we fix  $b^\mu = (0,0,0,L)$. 
\item The intermediate neutrinos are relativistic, hence everywhere we can approximate 
$\sqrt{\mathbf{q}^2+ m_k^2} \approx |\mathbf{q}| $.  
The  dependence on the neutrino masses survives only via the  $\Delta m_{kl}^2$ factor in \eref{NOF_tmp1}.
\item In \eref{NOF_tmp1} we can ignore the dependence on the mismatch momentum $\mathbf{r}$ everywhere except where it matters. 
More precisely, we keep $\mathbf{r}$ in   i)  the $e^{i b r} = e^{-i L r_z}$ factor (because it is multiplied by the large  $L$), 
 ii) the last denominator (because it goes through a pole), and iii) in $e^{i t}$ factor (in case $t$ is large).  
In the last two cases we keep terms up to linear order in  $\mathbf{r}$. 
On the other hand, we set $1/(E_S' E_T') = 1/(E_S m_T)$, and   $\cM' = \cM$. 
\een
With the above assumptions \eref{NOF_tmp1} simplifies to 
\beq
{d N_{\alpha \beta} \over d t}  =  {1 \over  64 \pi^3 E_S m_T} \sum_{kl} \int   d \Pi_P d \Pi_D  d^3 r   \cM_{\alpha k}^P \cM_{\beta k}^D 
 \bar \cM_{\alpha l}^{P} \bar \cM_{\beta l}^{D}
{ i  e^{-i L r_z} \over  2 \mathbf{q}\mathbf{r} - \Delta m_{kl}^2  + i \epsilon}  e^{i  {\bar p_S \mathbf{r} \over E_S} t }  + \hc  
\eeq 
We will assume for simplicity that $\bar p_S$ points along the source-target axis, $\bar p_S = (0,0,p)$.\footnote{%
Off-axis decays may be relevant in some experimental settings, and they can be similarly worked out using our approach.}
Then 
\beq
{d N_{\alpha \beta} \over d t} =   {1 \over 64 \pi^3 E_S m_T} \sum_{kl} \int   d \Pi_P d \Pi_D  d^3 r   \cM_{\alpha k}^P \cM_{\beta k}^D 
 \bar \cM_{\alpha l}^{P} \bar \cM_{\beta l}^{D}
{ i  e^{-i \bar L r_z} \over  2 \mathbf{q}\mathbf{r} - \Delta m_{kl}^2  + i \epsilon}   + \hc,  
\eeq 
where $\bar L = L - v t $, $v = p/E_S$. 
For concreteness we assume that $\bar L > 0$, and that the source does not live/travel long enough to make it  negative.  

We can now perform the integral over $r_z$, again treating it as a contour integral. 
The integrand has a single  pole at $r_z = (\Delta m_{kl}^2 - 2 q_i r_i  - i \epsilon)/(2 q_z)$ which falls 
below (above) the imaginary axis for $q_z$ positive (negative). 
Due to the $e^{-i \bar L r_z}$ factor in the numerator we are allowed to close the integration contour below but not above, 
and then we pick up the pole only  when $q_z > 0$. 
Thus we find  
\bea
{d N_{\alpha \beta} \over d t}  & = &  {1 \over 64 \pi^2 E_S m_T} \sum_{kl} \int   d \Pi_P d \Pi_D  d r_x d r_y   \cM_{\alpha k}^P \cM_{\beta k}^D 
 \bar \cM_{\alpha l}^{P} \bar \cM_{\beta l}^{D}
 \nnl &  \times & 
{\theta(q_z) \over q_z} \exp \big [ - i \bar L {\Delta m_{kl}^2- 2 q_x r_x - 2 q_y r_y 
 \over 2 q_z}    \big ]  + \hc  
\eea 
 The oscillatory factor $e^{- i 2 \pi \bar L/L_{\rm osc}}$
appears here  for the first time in this derivation, with the oscillation length  
$L_{\rm osc} =  {4 \pi q_z \over \Delta m_{kl}^2}$.  
In the QFT approach it arises because of interference between distinct neutrino mass eigenstates  $k \neq l$ entering the propagator, 
which in turn is possible due to the momentum mismatch $\mathbf{r}$ inherent in the description of initial wave packets. 
The integrals over $r_x$ and $r_y$ are now trivial,
and we arrive at 
\beq
 \label{eq:NOF_dPdt_generalized0}
{d N_{\alpha \beta}  \over dt d q_z }  
=
{\theta(q_z) q_z  \over 16  \bar L^2 E_S m_T } \sum_{kl}   
e^{-i {\bar L \Delta m_{kl}^2 \over 2 q_z}} 
 \int  d \Pi_{P'}   d \Pi_D    
 \cM_{\alpha k}^P \cM_{\beta k}^D \bar \cM_{\alpha l}^{P} \bar \cM_{\beta l}^{D} \delta(q_x) \delta(q_y)  + \hc, 
\eeq
where $ d \Pi_{P} =  d \Pi_{P'} d q_z$, and  $q_{x,y}$ ($q_z$) are the neutrino momenta orthogonal (parallel) to the source-target line.
Note that the dependence of the rate on the distance $\bar L$ between the source and the target enters in two ways: 
via the oscillatory $e^{- i 2 \pi \bar L/L_{\rm osc}}$ factor, and via the geometric $1/L^2$ factor. 
The latter can be intuitively understood as being due to the decrease of the overall neutrino flux with the increasing $L$. 
At this point it is  easy to see that the $\hc$ part returns the same expression after relabeling $k \leftrightarrow l$, thus we can simplify 
\beq
 \label{eq:NOF_dPdt_generalized}
{d N_{\alpha \beta}  \over dt d q_z }  
=
{\theta(q_z) q_z  \over 8  \bar L^2 E_S m_T } \sum_{kl}   
e^{-i {\bar L \Delta m_{kl}^2 \over 2 q_z}} 
 \int  d \Pi_{P'}   d \Pi_D    
 \cM_{\alpha k}^P \cM_{\beta k}^D \bar \cM_{\alpha l}^{P} \bar \cM_{\beta l}^{D} \delta(q_x) \delta(q_y) . 
\eeq
This is the final step of our derivation. 

\vspace{1cm}

\eref{NOF_dPdt_generalized} is quite general. 
It is valid for any single neutrino production or detection mechanism, any neutrino interactions, any number of light neutrinos, etc. 
It can be also used when the neutrino production is not isotropic, in particular, for neutrino produced via decays in flight, and for polarized neutrino production.
As discussed above \eref{NOF_tmp1}, one restriction for the validity of this oscillation formula  is 
\beq
\label{eq:NOF_oscondition}
\sigma \ll {q_z \over \Delta m_{kl}^2  } \sim L_{\rm osc}. 
\eeq 
When the packet size becomes comparable to the oscillation  length, then oscillations  are suppressed~\cite{Kayser:1981ye}.
In our derivation, this happens because, for $\sigma \gtrsim L_{\rm osc}$, the wave functions $f(p_j)$ and $f(p_j+r)^*$ no longer overlap for $|\mathbf{r}| \sim \Delta m_{kl}^2/q_z$.  
On the other hand, in our approach we do not find exponential suppression of the oscillations  proportional to the distance $L$ travelled by the neutrino.
The usual argument for this suppression~\cite{Nussinov:1976uw,Kayser:1981ye}, due the decoherence of wave packets corresponding to different neutrino eigenstates traveling at different speeds, does not apply in the stationary situation considered here.

For an easier comparison with the expressions from the earlier oscillation literature, we recast \eref{NOF_dPdt_generalized} in the case where  neutrinos are produced isotropically, that is $\cM_{\alpha k}^P \bar \cM_{\alpha l}^{P}$ integrate/summed over unobserved degrees of freedom is independent of the direction of the neutrino momentum.  
This assumption is satisfied in typical neutrino experiments where the source is  unpolarized.
Then one can first perform the (trivial) integral over $q_{x,y}$ to get rid of the delta function, and  then put back an (also trivial) integration over the neutrino angular phase space variables, 
$1 = \int {d \Omega_\nu \over 4 \pi}$. 
This leads to the expression 
\beq 
 \label{eq:NOF_dPdt}
{d N_{\alpha \beta}  \over dt d E_\nu }  
=
{1 \over  32 \pi \bar L^2 E_S m_T E_\nu } 
 \int  d \Pi_{P'}   d \Pi_D    \sum_{kl}  
 {e^{-i {\bar L \Delta m_{kl}^2 \over 2 E_\nu}}} 
 \cM_{\alpha k}^P \cM_{\beta k}^D \bar \cM_{\alpha l}^{P} \bar \cM_{\beta l}^{D} , 
\eeq
where $E_\nu = \theta(q_z) q_z$.
After multiplying the rate by the number of source and detector particles $N_{S,T}$, and choosing the source at rest, $E_S=m_S$, 
we obtain the master formula in \eref{rateqft}.

As a final comment, in this derivation we have assumed the absence of matter effects in propagation.  
The neutral-current NSI {\em other} than the matter effects can also be correctly described by QFT expressions analogous to \eref{NOF_dPdt_generalized},
and they are relevant e.g. if neutrinos are detected via coherent scattering on nuclei. 
To include NSI entering via the matter effects one would need to modify the neutrino propagator
starting from \eref{NOF_neutrinopropagator}.

\bibliographystyle{apsrev4-1}
\bibliography{nusQFT}

\begin{thebibliography}{50}%
\makeatletter
\providecommand \@ifxundefined [1]{%
 \@ifx{#1\undefined}
}%
\providecommand \@ifnum [1]{%
 \ifnum #1\expandafter \@firstoftwo
 \else \expandafter \@secondoftwo
 \fi
}%
\providecommand \@ifx [1]{%
 \ifx #1\expandafter \@firstoftwo
 \else \expandafter \@secondoftwo
 \fi
}%
\providecommand \natexlab [1]{#1}%
\providecommand \enquote  [1]{``#1''}%
\providecommand \bibnamefont  [1]{#1}%
\providecommand \bibfnamefont [1]{#1}%
\providecommand \citenamefont [1]{#1}%
\providecommand \href@noop [0]{\@secondoftwo}%
\providecommand \href [0]{\begingroup \@sanitize@url \@href}%
\providecommand \@href[1]{\@@startlink{#1}\@@href}%
\providecommand \@@href[1]{\endgroup#1\@@endlink}%
\providecommand \@sanitize@url [0]{\catcode `\\12\catcode `\$12\catcode
  `\&12\catcode `\#12\catcode `\^12\catcode `\_12\catcode `\%12\relax}%
\providecommand \@@startlink[1]{}%
\providecommand \@@endlink[0]{}%
\providecommand \url  [0]{\begingroup\@sanitize@url \@url }%
\providecommand \@url [1]{\endgroup\@href {#1}{\urlprefix }}%
\providecommand \urlprefix  [0]{URL }%
\providecommand \Eprint [0]{\href }%
\providecommand \doibase [0]{http://dx.doi.org/}%
\providecommand \selectlanguage [0]{\@gobble}%
\providecommand \bibinfo  [0]{\@secondoftwo}%
\providecommand \bibfield  [0]{\@secondoftwo}%
\providecommand \translation [1]{[#1]}%
\providecommand \BibitemOpen [0]{}%
\providecommand \bibitemStop [0]{}%
\providecommand \bibitemNoStop [0]{.\EOS\space}%
\providecommand \EOS [0]{\spacefactor3000\relax}%
\providecommand \BibitemShut  [1]{\csname bibitem#1\endcsname}%
\let\auto@bib@innerbib\@empty
\bibitem [{\citenamefont {Bilenky}\ and\ \citenamefont
  {Pontecorvo}(1978)}]{Bilenky:1978nj}%
  \BibitemOpen
  \bibfield  {author} {\bibinfo {author} {\bibfnamefont {S.~M.}\ \bibnamefont
  {Bilenky}}\ and\ \bibinfo {author} {\bibfnamefont {B.}~\bibnamefont
  {Pontecorvo}},\ }\href {\doibase 10.1016/0370-1573(78)90095-9} {\bibfield
  {journal} {\bibinfo  {journal} {Phys. Rept.}\ }\textbf {\bibinfo {volume}
  {41}},\ \bibinfo {pages} {225} (\bibinfo {year} {1978})}\BibitemShut
  {NoStop}%
\bibitem [{\citenamefont {de~Salas}\ \emph {et~al.}(2018)\citenamefont
  {de~Salas}, \citenamefont {Forero}, \citenamefont {Ternes}, \citenamefont
  {Tortola},\ and\ \citenamefont {Valle}}]{deSalas:2017kay}%
  \BibitemOpen
  \bibfield  {author} {\bibinfo {author} {\bibfnamefont {P.~F.}\ \bibnamefont
  {de~Salas}}, \bibinfo {author} {\bibfnamefont {D.~V.}\ \bibnamefont
  {Forero}}, \bibinfo {author} {\bibfnamefont {C.~A.}\ \bibnamefont {Ternes}},
  \bibinfo {author} {\bibfnamefont {M.}~\bibnamefont {Tortola}}, \ and\
  \bibinfo {author} {\bibfnamefont {J.~W.~F.}\ \bibnamefont {Valle}},\ }\href
  {\doibase 10.1016/j.physletb.2018.06.019} {\bibfield  {journal} {\bibinfo
  {journal} {Phys. Lett.}\ }\textbf {\bibinfo {volume} {B782}},\ \bibinfo
  {pages} {633} (\bibinfo {year} {2018})},\ \Eprint
  {http://arxiv.org/abs/1708.01186} {arXiv:1708.01186 [hep-ph]} \BibitemShut
  {NoStop}%
\bibitem [{\citenamefont {Esteban}\ \emph {et~al.}(2019)\citenamefont
  {Esteban}, \citenamefont {Gonzalez-Garcia}, \citenamefont
  {Hernandez-Cabezudo}, \citenamefont {Maltoni},\ and\ \citenamefont
  {Schwetz}}]{Esteban:2018azc}%
  \BibitemOpen
  \bibfield  {author} {\bibinfo {author} {\bibfnamefont {I.}~\bibnamefont
  {Esteban}}, \bibinfo {author} {\bibfnamefont {M.~C.}\ \bibnamefont
  {Gonzalez-Garcia}}, \bibinfo {author} {\bibfnamefont {A.}~\bibnamefont
  {Hernandez-Cabezudo}}, \bibinfo {author} {\bibfnamefont {M.}~\bibnamefont
  {Maltoni}}, \ and\ \bibinfo {author} {\bibfnamefont {T.}~\bibnamefont
  {Schwetz}},\ }\href {\doibase 10.1007/JHEP01(2019)106} {\bibfield  {journal}
  {\bibinfo  {journal} {JHEP}\ }\textbf {\bibinfo {volume} {01}},\ \bibinfo
  {pages} {106} (\bibinfo {year} {2019})},\ \Eprint
  {http://arxiv.org/abs/1811.05487} {arXiv:1811.05487 [hep-ph]} \BibitemShut
  {NoStop}%
\bibitem [{\citenamefont {Weinberg}(1979)}]{Weinberg:1979sa}%
  \BibitemOpen
  \bibfield  {author} {\bibinfo {author} {\bibfnamefont {S.}~\bibnamefont
  {Weinberg}},\ }\href {\doibase 10.1103/PhysRevLett.43.1566} {\bibfield
  {journal} {\bibinfo  {journal} {Phys. Rev. Lett.}\ }\textbf {\bibinfo
  {volume} {43}},\ \bibinfo {pages} {1566} (\bibinfo {year}
  {1979})}\BibitemShut {NoStop}%
\bibitem [{\citenamefont {Bergmann}\ \emph {et~al.}(1999)\citenamefont
  {Bergmann}, \citenamefont {Grossman},\ and\ \citenamefont
  {Nardi}}]{Bergmann:1999rz}%
  \BibitemOpen
  \bibfield  {author} {\bibinfo {author} {\bibfnamefont {S.}~\bibnamefont
  {Bergmann}}, \bibinfo {author} {\bibfnamefont {Y.}~\bibnamefont {Grossman}},
  \ and\ \bibinfo {author} {\bibfnamefont {E.}~\bibnamefont {Nardi}},\ }\href
  {\doibase 10.1103/PhysRevD.60.093008} {\bibfield  {journal} {\bibinfo
  {journal} {Phys. Rev.}\ }\textbf {\bibinfo {volume} {D60}},\ \bibinfo {pages}
  {093008} (\bibinfo {year} {1999})},\ \Eprint
  {http://arxiv.org/abs/hep-ph/9903517} {arXiv:hep-ph/9903517 [hep-ph]}
  \BibitemShut {NoStop}%
\bibitem [{\citenamefont {Antusch}\ \emph {et~al.}(2006)\citenamefont
  {Antusch}, \citenamefont {Biggio}, \citenamefont {Fernandez-Martinez},
  \citenamefont {Gavela},\ and\ \citenamefont {Lopez-Pavon}}]{Antusch:2006vwa}%
  \BibitemOpen
  \bibfield  {author} {\bibinfo {author} {\bibfnamefont {S.}~\bibnamefont
  {Antusch}}, \bibinfo {author} {\bibfnamefont {C.}~\bibnamefont {Biggio}},
  \bibinfo {author} {\bibfnamefont {E.}~\bibnamefont {Fernandez-Martinez}},
  \bibinfo {author} {\bibfnamefont {M.~B.}\ \bibnamefont {Gavela}}, \ and\
  \bibinfo {author} {\bibfnamefont {J.}~\bibnamefont {Lopez-Pavon}},\ }\href
  {\doibase 10.1088/1126-6708/2006/10/084} {\bibfield  {journal} {\bibinfo
  {journal} {JHEP}\ }\textbf {\bibinfo {volume} {10}},\ \bibinfo {pages} {084}
  (\bibinfo {year} {2006})},\ \Eprint {http://arxiv.org/abs/hep-ph/0607020}
  {arXiv:hep-ph/0607020 [hep-ph]} \BibitemShut {NoStop}%
\bibitem [{\citenamefont {Kopp}\ \emph {et~al.}(2008)\citenamefont {Kopp},
  \citenamefont {Lindner}, \citenamefont {Ota},\ and\ \citenamefont
  {Sato}}]{Kopp:2007ne}%
  \BibitemOpen
  \bibfield  {author} {\bibinfo {author} {\bibfnamefont {J.}~\bibnamefont
  {Kopp}}, \bibinfo {author} {\bibfnamefont {M.}~\bibnamefont {Lindner}},
  \bibinfo {author} {\bibfnamefont {T.}~\bibnamefont {Ota}}, \ and\ \bibinfo
  {author} {\bibfnamefont {J.}~\bibnamefont {Sato}},\ }\href {\doibase
  10.1103/PhysRevD.77.013007} {\bibfield  {journal} {\bibinfo  {journal} {Phys.
  Rev.}\ }\textbf {\bibinfo {volume} {D77}},\ \bibinfo {pages} {013007}
  (\bibinfo {year} {2008})},\ \Eprint {http://arxiv.org/abs/0708.0152}
  {arXiv:0708.0152 [hep-ph]} \BibitemShut {NoStop}%
\bibitem [{\citenamefont {Bolanos}\ \emph {et~al.}(2009)\citenamefont
  {Bolanos}, \citenamefont {Miranda}, \citenamefont {Palazzo}, \citenamefont
  {Tortola},\ and\ \citenamefont {Valle}}]{Bolanos:2008km}%
  \BibitemOpen
  \bibfield  {author} {\bibinfo {author} {\bibfnamefont {A.}~\bibnamefont
  {Bolanos}}, \bibinfo {author} {\bibfnamefont {O.~G.}\ \bibnamefont
  {Miranda}}, \bibinfo {author} {\bibfnamefont {A.}~\bibnamefont {Palazzo}},
  \bibinfo {author} {\bibfnamefont {M.~A.}\ \bibnamefont {Tortola}}, \ and\
  \bibinfo {author} {\bibfnamefont {J.~W.~F.}\ \bibnamefont {Valle}},\ }\href
  {\doibase 10.1103/PhysRevD.79.113012} {\bibfield  {journal} {\bibinfo
  {journal} {Phys. Rev.}\ }\textbf {\bibinfo {volume} {D79}},\ \bibinfo {pages}
  {113012} (\bibinfo {year} {2009})},\ \Eprint {http://arxiv.org/abs/0812.4417}
  {arXiv:0812.4417 [hep-ph]} \BibitemShut {NoStop}%
\bibitem [{\citenamefont {Ohlsson}\ and\ \citenamefont
  {Zhang}(2009)}]{Ohlsson:2008gx}%
  \BibitemOpen
  \bibfield  {author} {\bibinfo {author} {\bibfnamefont {T.}~\bibnamefont
  {Ohlsson}}\ and\ \bibinfo {author} {\bibfnamefont {H.}~\bibnamefont
  {Zhang}},\ }\href {\doibase 10.1016/j.physletb.2008.12.005} {\bibfield
  {journal} {\bibinfo  {journal} {Phys. Lett.}\ }\textbf {\bibinfo {volume}
  {B671}},\ \bibinfo {pages} {99} (\bibinfo {year} {2009})},\ \Eprint
  {http://arxiv.org/abs/0809.4835} {arXiv:0809.4835 [hep-ph]} \BibitemShut
  {NoStop}%
\bibitem [{\citenamefont {Delepine}\ \emph {et~al.}(2009)\citenamefont
  {Delepine}, \citenamefont {Gonzalez~Macias}, \citenamefont {Khalil},\ and\
  \citenamefont {Lopez~Castro}}]{Delepine:2009am}%
  \BibitemOpen
  \bibfield  {author} {\bibinfo {author} {\bibfnamefont {D.}~\bibnamefont
  {Delepine}}, \bibinfo {author} {\bibfnamefont {V.}~\bibnamefont
  {Gonzalez~Macias}}, \bibinfo {author} {\bibfnamefont {S.}~\bibnamefont
  {Khalil}}, \ and\ \bibinfo {author} {\bibfnamefont {G.}~\bibnamefont
  {Lopez~Castro}},\ }\href {\doibase 10.1103/PhysRevD.79.093003} {\bibfield
  {journal} {\bibinfo  {journal} {Phys. Rev.}\ }\textbf {\bibinfo {volume}
  {D79}},\ \bibinfo {pages} {093003} (\bibinfo {year} {2009})},\ \Eprint
  {http://arxiv.org/abs/0901.1460} {arXiv:0901.1460 [hep-ph]} \BibitemShut
  {NoStop}%
\bibitem [{\citenamefont {Biggio}\ \emph {et~al.}(2009)\citenamefont {Biggio},
  \citenamefont {Blennow},\ and\ \citenamefont
  {Fernandez-Martinez}}]{Biggio:2009nt}%
  \BibitemOpen
  \bibfield  {author} {\bibinfo {author} {\bibfnamefont {C.}~\bibnamefont
  {Biggio}}, \bibinfo {author} {\bibfnamefont {M.}~\bibnamefont {Blennow}}, \
  and\ \bibinfo {author} {\bibfnamefont {E.}~\bibnamefont
  {Fernandez-Martinez}},\ }\href {\doibase 10.1088/1126-6708/2009/08/090}
  {\bibfield  {journal} {\bibinfo  {journal} {JHEP}\ }\textbf {\bibinfo
  {volume} {08}},\ \bibinfo {pages} {090} (\bibinfo {year} {2009})},\ \Eprint
  {http://arxiv.org/abs/0907.0097} {arXiv:0907.0097 [hep-ph]} \BibitemShut
  {NoStop}%
\bibitem [{\citenamefont {Leitner}\ \emph {et~al.}(2011)\citenamefont
  {Leitner}, \citenamefont {Malinsky}, \citenamefont {Roskovec},\ and\
  \citenamefont {Zhang}}]{Leitner:2011aa}%
  \BibitemOpen
  \bibfield  {author} {\bibinfo {author} {\bibfnamefont {R.}~\bibnamefont
  {Leitner}}, \bibinfo {author} {\bibfnamefont {M.}~\bibnamefont {Malinsky}},
  \bibinfo {author} {\bibfnamefont {B.}~\bibnamefont {Roskovec}}, \ and\
  \bibinfo {author} {\bibfnamefont {H.}~\bibnamefont {Zhang}},\ }\href
  {\doibase 10.1007/JHEP12(2011)001} {\bibfield  {journal} {\bibinfo  {journal}
  {JHEP}\ }\textbf {\bibinfo {volume} {12}},\ \bibinfo {pages} {001} (\bibinfo
  {year} {2011})},\ \Eprint {http://arxiv.org/abs/1105.5580} {arXiv:1105.5580
  [hep-ph]} \BibitemShut {NoStop}%
\bibitem [{\citenamefont {Ohlsson}(2013)}]{Ohlsson:2012kf}%
  \BibitemOpen
  \bibfield  {author} {\bibinfo {author} {\bibfnamefont {T.}~\bibnamefont
  {Ohlsson}},\ }\href {\doibase 10.1088/0034-4885/76/4/044201} {\bibfield
  {journal} {\bibinfo  {journal} {Rept. Prog. Phys.}\ }\textbf {\bibinfo
  {volume} {76}},\ \bibinfo {pages} {044201} (\bibinfo {year} {2013})},\
  \Eprint {http://arxiv.org/abs/1209.2710} {arXiv:1209.2710 [hep-ph]}
  \BibitemShut {NoStop}%
\bibitem [{\citenamefont {Esmaili}\ and\ \citenamefont
  {Smirnov}(2013)}]{Esmaili:2013fva}%
  \BibitemOpen
  \bibfield  {author} {\bibinfo {author} {\bibfnamefont {A.}~\bibnamefont
  {Esmaili}}\ and\ \bibinfo {author} {\bibfnamefont {A.~{\relax Yu}.}\
  \bibnamefont {Smirnov}},\ }\href {\doibase 10.1007/JHEP06(2013)026}
  {\bibfield  {journal} {\bibinfo  {journal} {JHEP}\ }\textbf {\bibinfo
  {volume} {06}},\ \bibinfo {pages} {026} (\bibinfo {year} {2013})},\ \Eprint
  {http://arxiv.org/abs/1304.1042} {arXiv:1304.1042 [hep-ph]} \BibitemShut
  {NoStop}%
\bibitem [{\citenamefont {Li}\ and\ \citenamefont {Zhou}(2014)}]{Li:2014mlo}%
  \BibitemOpen
  \bibfield  {author} {\bibinfo {author} {\bibfnamefont {Y.-F.}\ \bibnamefont
  {Li}}\ and\ \bibinfo {author} {\bibfnamefont {Y.-L.}\ \bibnamefont {Zhou}},\
  }\href {\doibase 10.1016/j.nuclphysb.2014.09.013} {\bibfield  {journal}
  {\bibinfo  {journal} {Nucl. Phys.}\ }\textbf {\bibinfo {volume} {B888}},\
  \bibinfo {pages} {137} (\bibinfo {year} {2014})},\ \Eprint
  {http://arxiv.org/abs/1408.6301} {arXiv:1408.6301 [hep-ph]} \BibitemShut
  {NoStop}%
\bibitem [{\citenamefont {Agarwalla}\ \emph {et~al.}(2015)\citenamefont
  {Agarwalla}, \citenamefont {Bagchi}, \citenamefont {Forero},\ and\
  \citenamefont {Tortola}}]{Agarwalla:2014bsa}%
  \BibitemOpen
  \bibfield  {author} {\bibinfo {author} {\bibfnamefont {S.~K.}\ \bibnamefont
  {Agarwalla}}, \bibinfo {author} {\bibfnamefont {P.}~\bibnamefont {Bagchi}},
  \bibinfo {author} {\bibfnamefont {D.~V.}\ \bibnamefont {Forero}}, \ and\
  \bibinfo {author} {\bibfnamefont {M.}~\bibnamefont {Tortola}},\ }\href
  {\doibase 10.1007/JHEP07(2015)060} {\bibfield  {journal} {\bibinfo  {journal}
  {JHEP}\ }\textbf {\bibinfo {volume} {07}},\ \bibinfo {pages} {060} (\bibinfo
  {year} {2015})},\ \Eprint {http://arxiv.org/abs/1412.1064} {arXiv:1412.1064
  [hep-ph]} \BibitemShut {NoStop}%
\bibitem [{\citenamefont {Blennow}\ \emph {et~al.}(2015)\citenamefont
  {Blennow}, \citenamefont {Choubey}, \citenamefont {Ohlsson},\ and\
  \citenamefont {Raut}}]{Blennow:2015nxa}%
  \BibitemOpen
  \bibfield  {author} {\bibinfo {author} {\bibfnamefont {M.}~\bibnamefont
  {Blennow}}, \bibinfo {author} {\bibfnamefont {S.}~\bibnamefont {Choubey}},
  \bibinfo {author} {\bibfnamefont {T.}~\bibnamefont {Ohlsson}}, \ and\
  \bibinfo {author} {\bibfnamefont {S.~K.}\ \bibnamefont {Raut}},\ }\href
  {\doibase 10.1007/JHEP09(2015)096} {\bibfield  {journal} {\bibinfo  {journal}
  {JHEP}\ }\textbf {\bibinfo {volume} {09}},\ \bibinfo {pages} {096} (\bibinfo
  {year} {2015})},\ \Eprint {http://arxiv.org/abs/1507.02868} {arXiv:1507.02868
  [hep-ph]} \BibitemShut {NoStop}%
\bibitem [{\citenamefont {Coloma}\ \emph {et~al.}(2017)\citenamefont {Coloma},
  \citenamefont {Gonzalez-Garcia}, \citenamefont {Maltoni},\ and\ \citenamefont
  {Schwetz}}]{Coloma:2017ncl}%
  \BibitemOpen
  \bibfield  {author} {\bibinfo {author} {\bibfnamefont {P.}~\bibnamefont
  {Coloma}}, \bibinfo {author} {\bibfnamefont {M.~C.}\ \bibnamefont
  {Gonzalez-Garcia}}, \bibinfo {author} {\bibfnamefont {M.}~\bibnamefont
  {Maltoni}}, \ and\ \bibinfo {author} {\bibfnamefont {T.}~\bibnamefont
  {Schwetz}},\ }\href {\doibase 10.1103/PhysRevD.96.115007} {\bibfield
  {journal} {\bibinfo  {journal} {Phys. Rev.}\ }\textbf {\bibinfo {volume}
  {D96}},\ \bibinfo {pages} {115007} (\bibinfo {year} {2017})},\ \Eprint
  {http://arxiv.org/abs/1708.02899} {arXiv:1708.02899 [hep-ph]} \BibitemShut
  {NoStop}%
\bibitem [{\citenamefont {Farzan}\ and\ \citenamefont
  {Tortola}(2018)}]{Farzan:2017xzy}%
  \BibitemOpen
  \bibfield  {author} {\bibinfo {author} {\bibfnamefont {Y.}~\bibnamefont
  {Farzan}}\ and\ \bibinfo {author} {\bibfnamefont {M.}~\bibnamefont
  {Tortola}},\ }\href {\doibase 10.3389/fphy.2018.00010} {\bibfield  {journal}
  {\bibinfo  {journal} {Front.in Phys.}\ }\textbf {\bibinfo {volume} {6}},\
  \bibinfo {pages} {10} (\bibinfo {year} {2018})},\ \Eprint
  {http://arxiv.org/abs/1710.09360} {arXiv:1710.09360 [hep-ph]} \BibitemShut
  {NoStop}%
\bibitem [{\citenamefont {Choudhury}\ \emph {et~al.}(2018)\citenamefont
  {Choudhury}, \citenamefont {Ghosh},\ and\ \citenamefont
  {Niyogi}}]{Choudhury:2018xsm}%
  \BibitemOpen
  \bibfield  {author} {\bibinfo {author} {\bibfnamefont {D.}~\bibnamefont
  {Choudhury}}, \bibinfo {author} {\bibfnamefont {K.}~\bibnamefont {Ghosh}}, \
  and\ \bibinfo {author} {\bibfnamefont {S.}~\bibnamefont {Niyogi}},\ }\href
  {\doibase 10.1016/j.physletb.2018.07.053} {\bibfield  {journal} {\bibinfo
  {journal} {Phys. Lett. B}\ }\textbf {\bibinfo {volume} {784}},\ \bibinfo
  {pages} {248} (\bibinfo {year} {2018})},\ \Eprint
  {http://arxiv.org/abs/1801.01513} {arXiv:1801.01513 [hep-ph]} \BibitemShut
  {NoStop}%
\bibitem [{\citenamefont {Heeck}\ \emph {et~al.}(2019)\citenamefont {Heeck},
  \citenamefont {Lindner}, \citenamefont {Rodejohann},\ and\ \citenamefont
  {Vogl}}]{Heeck:2018nzc}%
  \BibitemOpen
  \bibfield  {author} {\bibinfo {author} {\bibfnamefont {J.}~\bibnamefont
  {Heeck}}, \bibinfo {author} {\bibfnamefont {M.}~\bibnamefont {Lindner}},
  \bibinfo {author} {\bibfnamefont {W.}~\bibnamefont {Rodejohann}}, \ and\
  \bibinfo {author} {\bibfnamefont {S.}~\bibnamefont {Vogl}},\ }\href {\doibase
  10.21468/SciPostPhys.6.3.038} {\bibfield  {journal} {\bibinfo  {journal}
  {SciPost Phys.}\ }\textbf {\bibinfo {volume} {6}},\ \bibinfo {pages} {038}
  (\bibinfo {year} {2019})},\ \Eprint {http://arxiv.org/abs/1812.04067}
  {arXiv:1812.04067 [hep-ph]} \BibitemShut {NoStop}%
\bibitem [{\citenamefont {Altmannshofer}\ \emph {et~al.}(2019)\citenamefont
  {Altmannshofer}, \citenamefont {Tammaro},\ and\ \citenamefont
  {Zupan}}]{Altmannshofer:2018xyo}%
  \BibitemOpen
  \bibfield  {author} {\bibinfo {author} {\bibfnamefont {W.}~\bibnamefont
  {Altmannshofer}}, \bibinfo {author} {\bibfnamefont {M.}~\bibnamefont
  {Tammaro}}, \ and\ \bibinfo {author} {\bibfnamefont {J.}~\bibnamefont
  {Zupan}},\ }\href {\doibase 10.1007/JHEP09(2019)083} {\bibfield  {journal}
  {\bibinfo  {journal} {JHEP}\ }\textbf {\bibinfo {volume} {09}},\ \bibinfo
  {pages} {083} (\bibinfo {year} {2019})},\ \Eprint
  {http://arxiv.org/abs/1812.02778} {arXiv:1812.02778 [hep-ph]} \BibitemShut
  {NoStop}%
\bibitem [{\citenamefont {Aristizabal~Sierra}\ \emph
  {et~al.}(2018)\citenamefont {Aristizabal~Sierra}, \citenamefont {De~Romeri},\
  and\ \citenamefont {Rojas}}]{AristizabalSierra:2018eqm}%
  \BibitemOpen
  \bibfield  {author} {\bibinfo {author} {\bibfnamefont {D.}~\bibnamefont
  {Aristizabal~Sierra}}, \bibinfo {author} {\bibfnamefont {V.}~\bibnamefont
  {De~Romeri}}, \ and\ \bibinfo {author} {\bibfnamefont {N.}~\bibnamefont
  {Rojas}},\ }\href {\doibase 10.1103/PhysRevD.98.075018} {\bibfield  {journal}
  {\bibinfo  {journal} {Phys. Rev.}\ }\textbf {\bibinfo {volume} {D98}},\
  \bibinfo {pages} {075018} (\bibinfo {year} {2018})},\ \Eprint
  {http://arxiv.org/abs/1806.07424} {arXiv:1806.07424 [hep-ph]} \BibitemShut
  {NoStop}%
\bibitem [{\citenamefont {Esteban}\ \emph {et~al.}(2018)\citenamefont
  {Esteban}, \citenamefont {Gonzalez-Garcia}, \citenamefont {Maltoni},
  \citenamefont {Martinez-Soler},\ and\ \citenamefont
  {Salvado}}]{Esteban:2018ppq}%
  \BibitemOpen
  \bibfield  {author} {\bibinfo {author} {\bibfnamefont {I.}~\bibnamefont
  {Esteban}}, \bibinfo {author} {\bibfnamefont {M.~C.}\ \bibnamefont
  {Gonzalez-Garcia}}, \bibinfo {author} {\bibfnamefont {M.}~\bibnamefont
  {Maltoni}}, \bibinfo {author} {\bibfnamefont {I.}~\bibnamefont
  {Martinez-Soler}}, \ and\ \bibinfo {author} {\bibfnamefont {J.}~\bibnamefont
  {Salvado}},\ }\href {\doibase 10.1007/JHEP08(2018)180} {\bibfield  {journal}
  {\bibinfo  {journal} {JHEP}\ }\textbf {\bibinfo {volume} {08}},\ \bibinfo
  {pages} {180} (\bibinfo {year} {2018})},\ \Eprint
  {http://arxiv.org/abs/1805.04530} {arXiv:1805.04530 [hep-ph]} \BibitemShut
  {NoStop}%
\bibitem [{\citenamefont {Giunti}\ \emph {et~al.}(1993)\citenamefont {Giunti},
  \citenamefont {Kim}, \citenamefont {Lee},\ and\ \citenamefont
  {Lee}}]{Giunti:1993se}%
  \BibitemOpen
  \bibfield  {author} {\bibinfo {author} {\bibfnamefont {C.}~\bibnamefont
  {Giunti}}, \bibinfo {author} {\bibfnamefont {C.~W.}\ \bibnamefont {Kim}},
  \bibinfo {author} {\bibfnamefont {J.~A.}\ \bibnamefont {Lee}}, \ and\
  \bibinfo {author} {\bibfnamefont {U.~W.}\ \bibnamefont {Lee}},\ }\href
  {\doibase 10.1103/PhysRevD.48.4310} {\bibfield  {journal} {\bibinfo
  {journal} {Phys. Rev.}\ }\textbf {\bibinfo {volume} {D48}},\ \bibinfo {pages}
  {4310} (\bibinfo {year} {1993})},\ \Eprint
  {http://arxiv.org/abs/hep-ph/9305276} {arXiv:hep-ph/9305276 [hep-ph]}
  \BibitemShut {NoStop}%
\bibitem [{\citenamefont {Akhmedov}\ and\ \citenamefont
  {Kopp}(2010)}]{Akhmedov:2010ms}%
  \BibitemOpen
  \bibfield  {author} {\bibinfo {author} {\bibfnamefont {E.~K.}\ \bibnamefont
  {Akhmedov}}\ and\ \bibinfo {author} {\bibfnamefont {J.}~\bibnamefont
  {Kopp}},\ }\href {\doibase 10.1007/JHEP04(2010)008, 10.1007/JHEP10(2013)052}
  {\bibfield  {journal} {\bibinfo  {journal} {JHEP}\ }\textbf {\bibinfo
  {volume} {04}},\ \bibinfo {pages} {008} (\bibinfo {year} {2010})},\ \bibinfo
  {note} {[Erratum: JHEP10,052(2013)]},\ \Eprint
  {http://arxiv.org/abs/1001.4815} {arXiv:1001.4815 [hep-ph]} \BibitemShut
  {NoStop}%
\bibitem [{\citenamefont {Kobach}\ \emph {et~al.}(2018)\citenamefont {Kobach},
  \citenamefont {Manohar},\ and\ \citenamefont {McGreevy}}]{Kobach:2017osm}%
  \BibitemOpen
  \bibfield  {author} {\bibinfo {author} {\bibfnamefont {A.}~\bibnamefont
  {Kobach}}, \bibinfo {author} {\bibfnamefont {A.~V.}\ \bibnamefont {Manohar}},
  \ and\ \bibinfo {author} {\bibfnamefont {J.}~\bibnamefont {McGreevy}},\
  }\href {\doibase 10.1016/j.physletb.2018.06.021} {\bibfield  {journal}
  {\bibinfo  {journal} {Phys. Lett.}\ }\textbf {\bibinfo {volume} {B783}},\
  \bibinfo {pages} {59} (\bibinfo {year} {2018})},\ \Eprint
  {http://arxiv.org/abs/1711.07491} {arXiv:1711.07491 [hep-ph]} \BibitemShut
  {NoStop}%
\bibitem [{\citenamefont {Giunti}\ and\ \citenamefont
  {Kim}(2007)}]{Giunti:2007ry}%
  \BibitemOpen
  \bibfield  {author} {\bibinfo {author} {\bibfnamefont {C.}~\bibnamefont
  {Giunti}}\ and\ \bibinfo {author} {\bibfnamefont {C.~W.}\ \bibnamefont
  {Kim}},\ }\href@noop {} {\emph {\bibinfo {title} {{Fundamentals of Neutrino
  Physics and Astrophysics}}}}\ (\bibinfo {year} {2007})\BibitemShut {NoStop}%
\bibitem [{\citenamefont {Grossman}(1995)}]{Grossman:1995wx}%
  \BibitemOpen
  \bibfield  {author} {\bibinfo {author} {\bibfnamefont {Y.}~\bibnamefont
  {Grossman}},\ }\href {\doibase 10.1016/0370-2693(95)01069-3} {\bibfield
  {journal} {\bibinfo  {journal} {Phys. Lett.}\ }\textbf {\bibinfo {volume}
  {B359}},\ \bibinfo {pages} {141} (\bibinfo {year} {1995})},\ \Eprint
  {http://arxiv.org/abs/hep-ph/9507344} {arXiv:hep-ph/9507344 [hep-ph]}
  \BibitemShut {NoStop}%
\bibitem [{\citenamefont {Gonzalez-Garcia}\ \emph {et~al.}(2001)\citenamefont
  {Gonzalez-Garcia}, \citenamefont {Grossman}, \citenamefont {Gusso},\ and\
  \citenamefont {Nir}}]{GonzalezGarcia:2001mp}%
  \BibitemOpen
  \bibfield  {author} {\bibinfo {author} {\bibfnamefont {M.~C.}\ \bibnamefont
  {Gonzalez-Garcia}}, \bibinfo {author} {\bibfnamefont {Y.}~\bibnamefont
  {Grossman}}, \bibinfo {author} {\bibfnamefont {A.}~\bibnamefont {Gusso}}, \
  and\ \bibinfo {author} {\bibfnamefont {Y.}~\bibnamefont {Nir}},\ }\href
  {\doibase 10.1103/PhysRevD.64.096006} {\bibfield  {journal} {\bibinfo
  {journal} {Phys. Rev.}\ }\textbf {\bibinfo {volume} {D64}},\ \bibinfo {pages}
  {096006} (\bibinfo {year} {2001})},\ \Eprint
  {http://arxiv.org/abs/hep-ph/0105159} {arXiv:hep-ph/0105159 [hep-ph]}
  \BibitemShut {NoStop}%
\bibitem [{\citenamefont {Cirigliano}\ \emph {et~al.}(2013)\citenamefont
  {Cirigliano}, \citenamefont {Gonzalez-Alonso},\ and\ \citenamefont
  {Graesser}}]{Cirigliano:2012ab}%
  \BibitemOpen
  \bibfield  {author} {\bibinfo {author} {\bibfnamefont {V.}~\bibnamefont
  {Cirigliano}}, \bibinfo {author} {\bibfnamefont {M.}~\bibnamefont
  {Gonzalez-Alonso}}, \ and\ \bibinfo {author} {\bibfnamefont {M.~L.}\
  \bibnamefont {Graesser}},\ }\href {\doibase 10.1007/JHEP02(2013)046}
  {\bibfield  {journal} {\bibinfo  {journal} {JHEP}\ }\textbf {\bibinfo
  {volume} {02}},\ \bibinfo {pages} {046} (\bibinfo {year} {2013})},\ \Eprint
  {http://arxiv.org/abs/1210.4553} {arXiv:1210.4553 [hep-ph]} \BibitemShut
  {NoStop}%
\bibitem [{\citenamefont {Falkowski}\ \emph {et~al.}(2019)\citenamefont
  {Falkowski}, \citenamefont {González-Alonso},\ and\ \citenamefont
  {Tabrizi}}]{Falkowski:2019xoe}%
  \BibitemOpen
  \bibfield  {author} {\bibinfo {author} {\bibfnamefont {A.}~\bibnamefont
  {Falkowski}}, \bibinfo {author} {\bibfnamefont {M.}~\bibnamefont
  {González-Alonso}}, \ and\ \bibinfo {author} {\bibfnamefont
  {Z.}~\bibnamefont {Tabrizi}},\ }\href {\doibase 10.1007/JHEP05(2019)173}
  {\bibfield  {journal} {\bibinfo  {journal} {JHEP}\ }\textbf {\bibinfo
  {volume} {05}},\ \bibinfo {pages} {173} (\bibinfo {year} {2019})},\ \Eprint
  {http://arxiv.org/abs/1901.04553} {arXiv:1901.04553 [hep-ph]} \BibitemShut
  {NoStop}%
\bibitem [{\citenamefont {Bischer}\ and\ \citenamefont
  {Rodejohann}(2019)}]{Bischer:2019ttk}%
  \BibitemOpen
  \bibfield  {author} {\bibinfo {author} {\bibfnamefont {I.}~\bibnamefont
  {Bischer}}\ and\ \bibinfo {author} {\bibfnamefont {W.}~\bibnamefont
  {Rodejohann}},\ }\href {\doibase 10.1016/j.nuclphysb.2019.114746} {\bibfield
  {journal} {\bibinfo  {journal} {Nucl. Phys.}\ }\textbf {\bibinfo {volume}
  {B947}},\ \bibinfo {pages} {114746} (\bibinfo {year} {2019})},\ \Eprint
  {http://arxiv.org/abs/1905.08699} {arXiv:1905.08699 [hep-ph]} \BibitemShut
  {NoStop}%
\bibitem [{\citenamefont {Khan}\ \emph {et~al.}(2013)\citenamefont {Khan},
  \citenamefont {McKay},\ and\ \citenamefont {Tahir}}]{Khan:2013hva}%
  \BibitemOpen
  \bibfield  {author} {\bibinfo {author} {\bibfnamefont {A.~N.}\ \bibnamefont
  {Khan}}, \bibinfo {author} {\bibfnamefont {D.~W.}\ \bibnamefont {McKay}}, \
  and\ \bibinfo {author} {\bibfnamefont {F.}~\bibnamefont {Tahir}},\ }\href
  {\doibase 10.1103/PhysRevD.88.113006} {\bibfield  {journal} {\bibinfo
  {journal} {Phys. Rev.}\ }\textbf {\bibinfo {volume} {D88}},\ \bibinfo {pages}
  {113006} (\bibinfo {year} {2013})},\ \Eprint {http://arxiv.org/abs/1305.4350}
  {arXiv:1305.4350 [hep-ph]} \BibitemShut {NoStop}%
\bibitem [{\citenamefont {Blennow}\ \emph {et~al.}(2017)\citenamefont
  {Blennow}, \citenamefont {Coloma}, \citenamefont {Fernandez-Martinez},
  \citenamefont {Hernandez-Garcia},\ and\ \citenamefont
  {Lopez-Pavon}}]{Blennow:2016jkn}%
  \BibitemOpen
  \bibfield  {author} {\bibinfo {author} {\bibfnamefont {M.}~\bibnamefont
  {Blennow}}, \bibinfo {author} {\bibfnamefont {P.}~\bibnamefont {Coloma}},
  \bibinfo {author} {\bibfnamefont {E.}~\bibnamefont {Fernandez-Martinez}},
  \bibinfo {author} {\bibfnamefont {J.}~\bibnamefont {Hernandez-Garcia}}, \
  and\ \bibinfo {author} {\bibfnamefont {J.}~\bibnamefont {Lopez-Pavon}},\
  }\href {\doibase 10.1007/JHEP04(2017)153} {\bibfield  {journal} {\bibinfo
  {journal} {JHEP}\ }\textbf {\bibinfo {volume} {04}},\ \bibinfo {pages} {153}
  (\bibinfo {year} {2017})},\ \Eprint {http://arxiv.org/abs/1609.08637}
  {arXiv:1609.08637 [hep-ph]} \BibitemShut {NoStop}%
\bibitem [{\citenamefont {Xing}(2012)}]{Xing:2011ur}%
  \BibitemOpen
  \bibfield  {author} {\bibinfo {author} {\bibfnamefont {Z.-z.}\ \bibnamefont
  {Xing}},\ }\href {\doibase 10.1103/PhysRevD.85.013008} {\bibfield  {journal}
  {\bibinfo  {journal} {Phys. Rev.}\ }\textbf {\bibinfo {volume} {D85}},\
  \bibinfo {pages} {013008} (\bibinfo {year} {2012})},\ \Eprint
  {http://arxiv.org/abs/1110.0083} {arXiv:1110.0083 [hep-ph]} \BibitemShut
  {NoStop}%
\bibitem [{\citenamefont {Cates}\ \emph {et~al.}(2011)\citenamefont {Cates},
  \citenamefont {de~Jager}, \citenamefont {Riordan},\ and\ \citenamefont
  {Wojtsekhowski}}]{Cates:2011pz}%
  \BibitemOpen
  \bibfield  {author} {\bibinfo {author} {\bibfnamefont {G.~D.}\ \bibnamefont
  {Cates}}, \bibinfo {author} {\bibfnamefont {C.~W.}\ \bibnamefont {de~Jager}},
  \bibinfo {author} {\bibfnamefont {S.}~\bibnamefont {Riordan}}, \ and\
  \bibinfo {author} {\bibfnamefont {B.}~\bibnamefont {Wojtsekhowski}},\ }\href
  {\doibase 10.1103/PhysRevLett.106.252003} {\bibfield  {journal} {\bibinfo
  {journal} {Phys. Rev. Lett.}\ }\textbf {\bibinfo {volume} {106}},\ \bibinfo
  {pages} {252003} (\bibinfo {year} {2011})},\ \Eprint
  {http://arxiv.org/abs/1103.1808} {arXiv:1103.1808 [nucl-ex]} \BibitemShut
  {NoStop}%
\bibitem [{\citenamefont {Gonzalez-Alonso}\ and\ \citenamefont
  {Martin~Camalich}(2014)}]{Gonzalez-Alonso:2013ura}%
  \BibitemOpen
  \bibfield  {author} {\bibinfo {author} {\bibfnamefont {M.}~\bibnamefont
  {Gonzalez-Alonso}}\ and\ \bibinfo {author} {\bibfnamefont {J.}~\bibnamefont
  {Martin~Camalich}},\ }\href {\doibase 10.1103/PhysRevLett.112.042501}
  {\bibfield  {journal} {\bibinfo  {journal} {Phys. Rev. Lett.}\ }\textbf
  {\bibinfo {volume} {112}},\ \bibinfo {pages} {042501} (\bibinfo {year}
  {2014})},\ \Eprint {http://arxiv.org/abs/1309.4434} {arXiv:1309.4434
  [hep-ph]} \BibitemShut {NoStop}%
\bibitem [{\citenamefont {Chang}\ \emph {et~al.}(2018)\citenamefont {Chang}
  \emph {et~al.}}]{Chang:2018uxx}%
  \BibitemOpen
  \bibfield  {author} {\bibinfo {author} {\bibfnamefont {C.~C.}\ \bibnamefont
  {Chang}} \emph {et~al.},\ }\href {\doibase 10.1038/s41586-018-0161-8}
  {\bibfield  {journal} {\bibinfo  {journal} {Nature}\ }\textbf {\bibinfo
  {volume} {558}},\ \bibinfo {pages} {91} (\bibinfo {year} {2018})},\ \Eprint
  {http://arxiv.org/abs/1805.12130} {arXiv:1805.12130 [hep-lat]} \BibitemShut
  {NoStop}%
\bibitem [{\citenamefont {Gupta}\ \emph {et~al.}(2018)\citenamefont {Gupta},
  \citenamefont {Jang}, \citenamefont {Yoon}, \citenamefont {Lin},
  \citenamefont {Cirigliano},\ and\ \citenamefont
  {Bhattacharya}}]{Gupta:2018qil}%
  \BibitemOpen
  \bibfield  {author} {\bibinfo {author} {\bibfnamefont {R.}~\bibnamefont
  {Gupta}}, \bibinfo {author} {\bibfnamefont {Y.-C.}\ \bibnamefont {Jang}},
  \bibinfo {author} {\bibfnamefont {B.}~\bibnamefont {Yoon}}, \bibinfo {author}
  {\bibfnamefont {H.-W.}\ \bibnamefont {Lin}}, \bibinfo {author} {\bibfnamefont
  {V.}~\bibnamefont {Cirigliano}}, \ and\ \bibinfo {author} {\bibfnamefont
  {T.}~\bibnamefont {Bhattacharya}},\ }\href {\doibase
  10.1103/PhysRevD.98.034503} {\bibfield  {journal} {\bibinfo  {journal} {Phys.
  Rev.}\ }\textbf {\bibinfo {volume} {D98}},\ \bibinfo {pages} {034503}
  (\bibinfo {year} {2018})},\ \Eprint {http://arxiv.org/abs/1806.09006}
  {arXiv:1806.09006 [hep-lat]} \BibitemShut {NoStop}%
\bibitem [{\citenamefont {Aoki}\ \emph {et~al.}(2020)\citenamefont {Aoki} \emph
  {et~al.}}]{Aoki:2019cca}%
  \BibitemOpen
  \bibfield  {author} {\bibinfo {author} {\bibfnamefont {S.}~\bibnamefont
  {Aoki}} \emph {et~al.} (\bibinfo {collaboration} {Flavour Lattice Averaging
  Group}),\ }\href {\doibase 10.1140/epjc/s10052-019-7354-7} {\bibfield
  {journal} {\bibinfo  {journal} {Eur. Phys. J. C}\ }\textbf {\bibinfo {volume}
  {80}},\ \bibinfo {pages} {113} (\bibinfo {year} {2020})},\ \Eprint
  {http://arxiv.org/abs/1902.08191} {arXiv:1902.08191 [hep-lat]} \BibitemShut
  {NoStop}%
\bibitem [{\citenamefont {Kopp}(2009)}]{Kopp:2009zza}%
  \BibitemOpen
  \bibfield  {author} {\bibinfo {author} {\bibfnamefont {J.}~\bibnamefont
  {Kopp}},\ }\emph {\bibinfo {title} {{New phenomena in neutrino physics}}},\
  \href {http://www.ub.uni-heidelberg.de/archiv/9381} {Ph.D. thesis},\ \bibinfo
   {school} {Heidelberg U.} (\bibinfo {year} {2009})\BibitemShut {NoStop}%
\bibitem [{\citenamefont {Langacker}\ and\ \citenamefont
  {London}(1988)}]{Langacker:1988up}%
  \BibitemOpen
  \bibfield  {author} {\bibinfo {author} {\bibfnamefont {P.}~\bibnamefont
  {Langacker}}\ and\ \bibinfo {author} {\bibfnamefont {D.}~\bibnamefont
  {London}},\ }\href {\doibase 10.1103/PhysRevD.38.907} {\bibfield  {journal}
  {\bibinfo  {journal} {Phys. Rev.}\ }\textbf {\bibinfo {volume} {D38}},\
  \bibinfo {pages} {907} (\bibinfo {year} {1988})}\BibitemShut {NoStop}%
\bibitem [{\citenamefont {Tang}\ and\ \citenamefont
  {Zhang}(2018)}]{Tang:2017qen}%
  \BibitemOpen
  \bibfield  {author} {\bibinfo {author} {\bibfnamefont {J.}~\bibnamefont
  {Tang}}\ and\ \bibinfo {author} {\bibfnamefont {Y.}~\bibnamefont {Zhang}},\
  }\href {\doibase 10.1103/PhysRevD.97.035018} {\bibfield  {journal} {\bibinfo
  {journal} {Phys. Rev.}\ }\textbf {\bibinfo {volume} {D97}},\ \bibinfo {pages}
  {035018} (\bibinfo {year} {2018})},\ \Eprint
  {http://arxiv.org/abs/1705.09500} {arXiv:1705.09500 [hep-ph]} \BibitemShut
  {NoStop}%
\bibitem [{\citenamefont {C.}\ \emph {et~al.}(2020)\citenamefont {C.},
  \citenamefont {Ghosh}, \citenamefont {Raut}, \citenamefont {Sinha},\ and\
  \citenamefont {Mehta}}]{C.:2019dbf}%
  \BibitemOpen
  \bibfield  {author} {\bibinfo {author} {\bibfnamefont {S.}~\bibnamefont
  {C.}}, \bibinfo {author} {\bibfnamefont {M.}~\bibnamefont {Ghosh}}, \bibinfo
  {author} {\bibfnamefont {S.~K.}\ \bibnamefont {Raut}}, \bibinfo {author}
  {\bibfnamefont {N.}~\bibnamefont {Sinha}}, \ and\ \bibinfo {author}
  {\bibfnamefont {P.}~\bibnamefont {Mehta}},\ }\href {\doibase
  10.1103/PhysRevD.101.055009} {\bibfield  {journal} {\bibinfo  {journal}
  {Phys. Rev.}\ }\textbf {\bibinfo {volume} {D101}},\ \bibinfo {pages} {055009}
  (\bibinfo {year} {2020})},\ \Eprint {http://arxiv.org/abs/1911.05021}
  {arXiv:1911.05021 [hep-ph]} \BibitemShut {NoStop}%
\bibitem [{\citenamefont {Descotes-Genon}\ \emph {et~al.}(2019)\citenamefont
  {Descotes-Genon}, \citenamefont {Falkowski}, \citenamefont {Fedele},
  \citenamefont {Gonzalez-Alonso},\ and\ \citenamefont
  {Virto}}]{Descotes-Genon:2018foz}%
  \BibitemOpen
  \bibfield  {author} {\bibinfo {author} {\bibfnamefont {S.}~\bibnamefont
  {Descotes-Genon}}, \bibinfo {author} {\bibfnamefont {A.}~\bibnamefont
  {Falkowski}}, \bibinfo {author} {\bibfnamefont {M.}~\bibnamefont {Fedele}},
  \bibinfo {author} {\bibfnamefont {M.}~\bibnamefont {Gonzalez-Alonso}}, \ and\
  \bibinfo {author} {\bibfnamefont {J.}~\bibnamefont {Virto}},\ }\href
  {\doibase 10.1007/JHEP05(2019)172} {\bibfield  {journal} {\bibinfo  {journal}
  {JHEP}\ }\textbf {\bibinfo {volume} {05}},\ \bibinfo {pages} {172} (\bibinfo
  {year} {2019})},\ \Eprint {http://arxiv.org/abs/1812.08163} {arXiv:1812.08163
  [hep-ph]} \BibitemShut {NoStop}%
\bibitem [{\citenamefont {Kosower}\ \emph {et~al.}(2019)\citenamefont
  {Kosower}, \citenamefont {Maybee},\ and\ \citenamefont
  {O'Connell}}]{Kosower:2018adc}%
  \BibitemOpen
  \bibfield  {author} {\bibinfo {author} {\bibfnamefont {D.~A.}\ \bibnamefont
  {Kosower}}, \bibinfo {author} {\bibfnamefont {B.}~\bibnamefont {Maybee}}, \
  and\ \bibinfo {author} {\bibfnamefont {D.}~\bibnamefont {O'Connell}},\ }\href
  {\doibase 10.1007/JHEP02(2019)137} {\bibfield  {journal} {\bibinfo  {journal}
  {JHEP}\ }\textbf {\bibinfo {volume} {02}},\ \bibinfo {pages} {137} (\bibinfo
  {year} {2019})},\ \Eprint {http://arxiv.org/abs/1811.10950} {arXiv:1811.10950
  [hep-th]} \BibitemShut {NoStop}%
\bibitem [{\citenamefont {Peskin}\ and\ \citenamefont
  {Schroeder}(1995)}]{Peskin:1995ev}%
  \BibitemOpen
  \bibfield  {author} {\bibinfo {author} {\bibfnamefont {M.~E.}\ \bibnamefont
  {Peskin}}\ and\ \bibinfo {author} {\bibfnamefont {D.~V.}\ \bibnamefont
  {Schroeder}},\ }\href {http://www.slac.stanford.edu/~mpeskin/QFT.html} {\emph
  {\bibinfo {title} {{An Introduction to quantum field theory}}}}\ (\bibinfo
  {publisher} {Addison-Wesley},\ \bibinfo {address} {Reading, USA},\ \bibinfo
  {year} {1995})\BibitemShut {NoStop}%
\bibitem [{\citenamefont {Kayser}(1981)}]{Kayser:1981ye}%
  \BibitemOpen
  \bibfield  {author} {\bibinfo {author} {\bibfnamefont {B.}~\bibnamefont
  {Kayser}},\ }\href {\doibase 10.1103/PhysRevD.24.110} {\bibfield  {journal}
  {\bibinfo  {journal} {Phys. Rev.}\ }\textbf {\bibinfo {volume} {D24}},\
  \bibinfo {pages} {110} (\bibinfo {year} {1981})}\BibitemShut {NoStop}%
\bibitem [{\citenamefont {Nussinov}(1976)}]{Nussinov:1976uw}%
  \BibitemOpen
  \bibfield  {author} {\bibinfo {author} {\bibfnamefont {S.}~\bibnamefont
  {Nussinov}},\ }\href {\doibase 10.1016/0370-2693(76)90648-1} {\bibfield
  {journal} {\bibinfo  {journal} {Phys. Lett.}\ }\textbf {\bibinfo {volume}
  {63B}},\ \bibinfo {pages} {201} (\bibinfo {year} {1976})}\BibitemShut
  {NoStop}%
\end{thebibliography}%

\end{document}